\documentclass[twocolumn,longauth]{aa}
\usepackage[utf8]{inputenc}

\usepackage{graphicx}
\usepackage{amsmath,amsfonts,amssymb}
\usepackage{txfonts}
\usepackage{color}
\usepackage{natbib}
\usepackage{float}
\usepackage{dblfloatfix}
\usepackage{afterpage}
\usepackage{ifthen}
\usepackage[morefloats=12]{morefloats}
\usepackage{placeins}
\usepackage{multicol}
\bibpunct{(}{)}{;}{a}{}{,}
\usepackage[switch]{lineno}
\definecolor{linkcolor}{rgb}{0.6,0,0}
\definecolor{citecolor}{rgb}{0,0,0.75}
\definecolor{urlcolor}{rgb}{0.12,0.46,0.7}
\usepackage[breaklinks, colorlinks, urlcolor=urlcolor,
    linkcolor=linkcolor,citecolor=citecolor,pdfencoding=auto]{hyperref}
\hypersetup{linktocpage}

\def\LiteBIRD{LiteBIRD} 
\def\litebird{LiteBIRD} 
 
\def\Planck{{\it Planck}} 
\def\WMAP{WMAP}

\def\FGBuster{\texttt{FGBuster}}

\def\Commander{\texttt{Commander}}
\def\commander{\texttt{Commander}}
\def\commanderone{\texttt{Commander1}}

\def\commanderthree{\texttt{Commander3}}

\def\NILC{\texttt{NILC}}
\def\nilc{\texttt{NILC}}
\def\cMILC{\texttt{cMILC}}
\def\cmilc{\texttt{cMILC}}

\renewcommand{\d}[0]{\vec{d}}

\newcommand{\red}[0]{\color{red}}

\newcommand{\s}[0]{\vec{s}}

\renewcommand{\L}[0]{\tens{L}}

\renewcommand{\r}[0]{\vec{r}}

\definecolor{gray}{RGB}{200,200,200}

\def\inv{^{-1}}

\def\setsymbol#1#2{\expandafter\def\csname #1\endcsname{#2}}
\def\getsymbol#1{\csname #1\endcsname}

\def\Planck{\textit{Planck}}

\newbox\tablebox    \newdimen\tablewidth
\def\leaderfil{\leaders\hbox to 5pt{\hss.\hss}\hfil}
\def\endPlancktablewide{\tablewidth=\textwidth 
    $$\hss\copy\tablebox\hss$$
    \vskip-\lastskip\vskip -2pt}
\def\tablenote#1 #2\par{\begingroup \parindent=0.8em
    \abovedisplayshortskip=0pt\belowdisplayshortskip=0pt
    \noindent
    $$\hss\vbox{\hsize\tablewidth \hangindent=\parindent \hangafter=1 \noindent
    \hbox to \parindent{$^#1$\hss}\strut#2\strut\par}\hss$$
    \endgroup}
\def\doubleline{\vskip 3pt\hrule \vskip 1.5pt \hrule \vskip 5pt}

\def\L2{\ifmmode L_2\else $L_2$\fi}

\def\DeltaT{\ifmmode \Delta T\else $\Delta T$\fi}
\def\deltat{\ifmmode \Delta t\else $\Delta t$\fi}
\def\fknee{\ifmmode f_{\rm knee}\else $f_{\rm knee}$\fi}
\def\Fmax{\ifmmode F_{\rm max}\else $F_{\rm max}$\fi}
\def\solar{\ifmmode{\rm M}_{\mathord\odot}\else${\rm M}_{\mathord\odot}$\fi}
\def\Msolar{\ifmmode{\rm M}_{\mathord\odot}\else${\rm M}_{\mathord\odot}$\fi}
\def\Lsolar{\ifmmode{\rm L}_{\mathord\odot}\else${\rm L}_{\mathord\odot}$\fi}
\def\inv{\ifmmode^{-1}\else$^{-1}$\fi}
\def\mo{\ifmmode^{-1}\else$^{-1}$\fi}
\def\sup#1{\ifmmode ^{\rm #1}\else $^{\rm #1}$\fi}
\def\expo#1{\ifmmode \times 10^{#1}\else $\times 10^{#1}$\fi}
\def\,{\thinspace}
\def\lsim{\mathrel{\raise .4ex\hbox{\rlap{$<$}\lower 1.2ex\hbox{$\sim$}}}}
\def\gsim{\mathrel{\raise .4ex\hbox{\rlap{$>$}\lower 1.2ex\hbox{$\sim$}}}}

\def\simprop{\mathrel{\raise .4ex\hbox{\rlap{$\propto$}\lower 1.2ex\hbox{$\sim$}}}}
\def\deg{\ifmmode^\circ\else$^\circ$\fi}
\def\pdeg{\ifmmode $\setbox0=\hbox{$^{\circ}$}\rlap{\hskip.11\wd0 .}$^{\circ}
          \else \setbox0=\hbox{$^{\circ}$}\rlap{\hskip.11\wd0 .}$^{\circ}$\fi}
\def\arcs{\ifmmode {^{\scriptstyle\prime\prime}}
          \else $^{\scriptstyle\prime\prime}$\fi}
\def\arcm{\ifmmode {^{\scriptstyle\prime}}
          \else $^{\scriptstyle\prime}$\fi}
\newdimen\sa  \newdimen\sb
\def\parcs{\sa=.07em \sb=.03em
     \ifmmode \hbox{\rlap{.}}^{\scriptstyle\prime\kern -\sb\prime}\hbox{\kern -\sa}
     \else \rlap{.}$^{\scriptstyle\prime\kern -\sb\prime}$\kern -\sa\fi}
\def\parcm{\sa=.08em \sb=.03em
     \ifmmode \hbox{\rlap{.}\kern\sa}^{\scriptstyle\prime}\hbox{\kern-\sb}
     \else \rlap{.}\kern\sa$^{\scriptstyle\prime}$\kern-\sb\fi}
\def\ra[#1 #2 #3.#4]{#1\sup{h}#2\sup{m}#3\sup{s}\llap.#4}
\def\dec[#1 #2 #3.#4]{#1\deg#2\arcm#3\arcs\llap.#4}
\def\deco[#1 #2 #3]{#1\deg#2\arcm#3\arcs}
\def\rra[#1 #2]{#1\sup{h}#2\sup{m}}

\def\dots{\relax\ifmmode \ldots\else $\ldots$\fi}
\def\WHzsr{\ifmmode $W\,Hz\mo\,sr\mo$\else W\,Hz\mo\,sr\mo\fi}
\def\mHz{\ifmmode $\,mHz$\else \,mHz\fi}
\def\GHz{\ifmmode $\,GHz$\else \,GHz\fi}
\def\mKs{\ifmmode $\,mK\,s$^{1/2}\else \,mK\,s$^{1/2}$\fi}
\def\muKs{\ifmmode \,\mu$K\,s$^{1/2}\else \,$\mu$K\,s$^{1/2}$\fi}
\def\muKRJs{\ifmmode \,\mu$K$_{\rm RJ}$\,s$^{1/2}\else \,$\mu$K$_{\rm RJ}$\,s$^{1/2}$\fi}
\def\muKHz{\ifmmode \,\mu$K\,Hz$^{-1/2}\else \,$\mu$K\,Hz$^{-1/2}$\fi}
\def\MJysr{\ifmmode \,$MJy\,sr\mo$\else \,MJy\,sr\mo\fi}
\def\MJysrmK{\ifmmode \,$MJy\,sr\mo$\,mK$_{\rm CMB}\mo\else \,MJy\,sr\mo\,mK$_{\rm CMB}\mo$\fi}
\def\microns{\ifmmode \,\mu$m$\else \,$\mu$m\fi}

\def\muK{\ifmmode \,\mu$K$\else \,$\mu$\hbox{K}\fi}
\def\microK{\ifmmode \,\mu$K$\else \,$\mu$\hbox{K}\fi}
\def\muW{\ifmmode \,\mu$W$\else \,$\mu$\hbox{W}\fi}
\def\kms{\ifmmode $\,km\,s$^{-1}\else \,km\,s$^{-1}$\fi}
\def\kmsMpc{\ifmmode $\,\kms\,Mpc\mo$\else \,\kms\,Mpc\mo\fi}
\providecommand{\sorthelp}[1]{}

\begin{document}

\title{Tensor-to-scalar ratio forecasts for \\extended LiteBIRD frequency configurations}
\author{\small
U.\,Fuskeland\inst{\ref{1}}\thanks{Corresponding author: U.~Fuskeland; \url{unnif@astro.uio.no}}\and
J.\,Aumont\inst{\ref{2}}\and
R.\,Aurlien\inst{\ref{1}}\and
C.\,Baccigalupi\inst{\ref{3},\ref{4},\ref{5}}\and
A.\,J.\,Banday\inst{\ref{2}}\and
H.\,K.\,Eriksen\inst{\ref{1}}\and
J.\,Errard\inst{\ref{6}}\and
R.\,T.\,Génova-Santos\inst{\ref{7},\ref{8}}\and
T.\,Hasebe\inst{\ref{9}}\and
J.\,Hubmayr\inst{\ref{10}}\and
H.\,Imada\inst{\ref{11}}\and
N.\,Krachmalnicoff\inst{\ref{3},\ref{4},\ref{5}}\and
L.\,Lamagna\inst{\ref{12},\ref{13}}\and
G.\,Pisano\inst{\ref{12}}\and
D.\,Poletti\inst{\ref{14},\ref{15}}\and
M.\,Remazeilles\inst{\ref{16},\ref{17}}\and
K.\,L.\,Thompson\inst{\ref{18},\ref{19}}\and
L.\,Vacher\inst{\ref{2}}\and
I.\,K.\,Wehus\inst{\ref{1}}\and
S.\,Azzoni\inst{\ref{20},\ref{9}}\and
M.\,Ballardini\inst{\ref{21},\ref{22},\ref{23}}\and
R.\,B.\,Barreiro\inst{\ref{16}}\and
N.\,Bartolo\inst{\ref{24},\ref{25},\ref{26}}\and
A.\,Basyrov\inst{\ref{1}}\and
D.\,Beck\inst{\ref{19}}\and
M.\,Bersanelli\inst{\ref{27},\ref{28}}\and
M.\,Bortolami\inst{\ref{21}}\and
M.\,Brilenkov\inst{\ref{1}}\and
E.\,Calabrese\inst{\ref{29}}\and
A.\,Carones\inst{\ref{30},\ref{31}}\and
F.\,J.\,Casas\inst{\ref{16}}\and
K.\,Cheung\inst{\ref{32},\ref{33},\ref{34}}\and
J.\,Chluba\inst{\ref{17}}\and
S.\,E.\,Clark\inst{\ref{19},\ref{35}}\and
L.\,Clermont\inst{\ref{36}}\and
F.\,Columbro\inst{\ref{12},\ref{13}}\and
A.\,Coppolecchia\inst{\ref{12},\ref{13}}\and
G.\,D'Alessandro\inst{\ref{12},\ref{13}}\and
P.\,de\,Bernardis\inst{\ref{12},\ref{13}}\and
T.\,de\,Haan\inst{\ref{37},\ref{38}}\and
E.\,de\,la\,Hoz\inst{\ref{16},\ref{39}}\and
M.\,De\,Petris\inst{\ref{12},\ref{13}}\and
S.\,Della\,Torre\inst{\ref{15}}\and
P.\,Diego-Palazuelos\inst{\ref{16},\ref{39}}\and
F.\,Finelli\inst{\ref{23},\ref{40}}\and
C.\,Franceschet\inst{\ref{27},\ref{28}}\and
G.\,Galloni\inst{\ref{30}}\and
M.\,Galloway\inst{\ref{1}}\and
M.\,Gerbino\inst{\ref{22}}\and
M.\,Gervasi\inst{\ref{14},\ref{15}}\and
T.\,Ghigna\inst{\ref{37}}\and
S.\,Giardiello\inst{\ref{29}}\and
E.\,Gjerløw\inst{\ref{1}}\and
A.\,Gruppuso\inst{\ref{23},\ref{40}}\and
P.\,Hargrave\inst{\ref{29}}\and
M.\,Hattori\inst{\ref{41}}\and
M.\,Hazumi\inst{\ref{38},\ref{37},\ref{42},\ref{9},\ref{43}}\and
L.\,T.\,Hergt\inst{\ref{44}}\and
D.\,Herman\inst{\ref{1}}\and
D.\,Herranz\inst{\ref{16}}\and
E.\,Hivon\inst{\ref{45}}\and
T.\,D.\,Hoang\inst{\ref{9}}\and
K.\,Kohri\inst{\ref{37}}\and
M.\,Lattanzi\inst{\ref{22}}\and
A.\,T.\,Lee\inst{\ref{46},\ref{32},\ref{38}}\and
C.\,Leloup\inst{\ref{6}}\and
F.\,Levrier\inst{\ref{47}}\and
A.\,I.\,Lonappan\inst{\ref{3}}\and
G.\,Luzzi\inst{\ref{48}}\and
B.\,Maffei\inst{\ref{49}}\and
E.\,Martínez-González\inst{\ref{16}}\and
S.\,Masi\inst{\ref{12},\ref{13}}\and
S.\,Matarrese\inst{\ref{24},\ref{25},\ref{26},\ref{50}}\and
T.\,Matsumura\inst{\ref{9}}\and
M.\,Migliaccio\inst{\ref{30},\ref{31}}\and
L.\,Montier\inst{\ref{2}}\and
G.\,Morgante\inst{\ref{23}}\and
B.\,Mot\inst{\ref{2}}\and
L.\,Mousset\inst{\ref{2}}\and
R.\,Nagata\inst{\ref{42}}\and
T.\,Namikawa\inst{\ref{9}}\and
F.\,Nati\inst{\ref{14},\ref{15}}\and
P.\,Natoli\inst{\ref{21},\ref{22}}\and
S.\,Nerval\inst{\ref{51}}\and
A.\,Novelli\inst{\ref{12}}\and
L.\,Pagano\inst{\ref{21},\ref{22},\ref{49}}\and
A.\,Paiella\inst{\ref{12},\ref{13}}\and
D.\,Paoletti\inst{\ref{23},\ref{40}}\and
G.\,Pascual-Cisneros\inst{\ref{16}}\and
G.\,Patanchon\inst{\ref{6}}\and
V.\,Pelgrims\inst{\ref{52},\ref{53}}\and
F.\,Piacentini\inst{\ref{12},\ref{13}}\and
G.\,Piccirilli\inst{\ref{30}}\and
G.\,Polenta\inst{\ref{48}}\and
G.\,Puglisi\inst{\ref{30},\ref{34}}\and
N.\,Raffuzzi\inst{\ref{21}}\and
A.\,Ritacco\inst{\ref{54},\ref{49},\ref{47}}\and
J.\,A.\,Rubino-Martin\inst{\ref{7},\ref{8}}\and
G.\,Savini\inst{\ref{55}}\and
D.\,Scott\inst{\ref{44}}\and
Y.\,Sekimoto\inst{\ref{42},\ref{56},\ref{37}}\and
M.\,Shiraishi\inst{\ref{57}}\and
G.\,Signorelli\inst{\ref{58}}\and
S.\,L.\,Stever\inst{\ref{59},\ref{9}}\and
N.\,Stutzer\inst{\ref{1}}\and
R.\,M.\,Sullivan\inst{\ref{44}}\and
H.\,Takakura\inst{\ref{56},\ref{42}}\and
L.\,Terenzi\inst{\ref{23}}\and
H.\,Thommesen\inst{\ref{1}}\and
M.\,Tristram\inst{\ref{60}}\and
M.\,Tsuji\inst{\ref{61}}\and
P.\,Vielva\inst{\ref{16}}\and
J.\,Weller\inst{\ref{62}}\and
B.\,Westbrook\inst{\ref{32}}\and
G.\,Weymann-Despres\inst{\ref{60}}\and
E.\,J.\,Wollack\inst{\ref{63}}\and
M.\,Zannoni\inst{\ref{14},\ref{15}}
\\LiteBIRD Collaboration.
}
\institute{\small
Institute of Theoretical Astrophysics, University of Oslo, Blindern, Oslo, Norway\label{1}\goodbreak\and
IRAP, Université de Toulouse, CNRS, CNES, UPS, (Toulouse), France\label{2}\goodbreak\and
International School for Advanced Studies (SISSA), Via Bonomea 265, 34136, Trieste, Italy\label{3}\goodbreak\and
INFN Sezione di Trieste, via Valerio 2, 34127 Trieste, Italy\label{4}\goodbreak\and
IFPU, Via Beirut, 2, 34151 Grignano, Trieste, Italy\label{5}\goodbreak\and
Université de Paris, CNRS, Astroparticule et Cosmologie, F-75013 Paris, France\label{6}\goodbreak\and
Instituto de Astrofísica de Canarias, E-38200 La Laguna, Tenerife, Canary Islands, Spain\label{7}\goodbreak\and
Departamento de Astrofísica, Universidad de La Laguna (ULL), E-38206, La Laguna, Tenerife, Spain\label{8}\goodbreak\and
Kavli Institute for the Physics and Mathematics of the Universe (Kavli IPMU, WPI), UTIAS, The University of Tokyo, Kashiwa, Chiba 277-8583, Japan\label{9}\goodbreak\and
NIST Quantum Sensors Group, 325 Broadway, Boulder, CO 80305, USA\label{10}\goodbreak\and
National Astronomical Observatory of Japan, Mitaka, Tokyo 181-8588, Japan\label{11}\goodbreak\and
Dipartimento di Fisica, Università La Sapienza, P. le A. Moro 2, Roma, Italy\label{12}\goodbreak\and
INFN Sezione di Roma, P.le A. Moro 2, 00185 Roma, Italy\label{13}\goodbreak\and
University of Milano Bicocca, Physics Department, p.zza della Scienza, 3, 20126 Milano, Italy\label{14}\goodbreak\and
INFN Sezione Milano Bicocca, Piazza della Scienza, 3, 20126 Milano, Italy\label{15}\goodbreak\and
Instituto de Fisica de Cantabria (IFCA, CSIC-UC), Avenida los Castros s/n, 39005, Santander, Spain\label{16}\goodbreak\and
Jodrell Bank Centre for Astrophysics, Department of Physics and Astronomy, School of Natural Sciences, The University of Manchester, Oxford Road, Manchester M13 9PL, UK\label{17}\goodbreak\and
SLAC National Accelerator Laboratory, Kavli Institute for Particle Astrophysics and Cosmology (KIPAC),  Menlo Park, CA 94025, USA\label{18}\goodbreak\and
Stanford University, Department of Physics,  CA 94305-4060, USA\label{19}\goodbreak\and
Department of Physics, University of Oxford, Denys Wilkinson Building, Keble Road, Oxford OX1 3RH, United Kingdom\label{20}\goodbreak\and
Dipartimento di Fisica e Scienze della Terra, Università di Ferrara, Via Saragat 1, 44122 Ferrara, Italy\label{21}\goodbreak\and
INFN Sezione di Ferrara, Via Saragat 1, 44122 Ferrara, Italy\label{22}\goodbreak\and
INAF - OAS Bologna, via Piero Gobetti, 93/3, 40129 Bologna, Italy\label{23}\goodbreak\and
Dipartimento di Fisica e Astronomia “G. Galilei”, Universita` degli Studi di Padova, via Marzolo 8, I-35131 Padova, Italy\label{24}\goodbreak\and
INFN Sezione di Padova, via Marzolo 8, I-35131, Padova, Italy\label{25}\goodbreak\and
INAF, Osservatorio Astronomico di Padova, Vicolo dell’Osservatorio 5, I-35122, Padova, Italy\label{26}\goodbreak\and
Dipartimento di Fisica, Universita' degli Studi di Milano, Via Celoria 16 - 20133, Milano, Italy\label{27}\goodbreak\and
INFN Sezione di Milano, Via Celoria 16 - 20133, Milano, Italy\label{28}\goodbreak\and
School of Physics and Astronomy, Cardiff University, Cardiff CF24 3AA, UK\label{29}\goodbreak\and
Dipartimento di Fisica, Università di Roma Tor Vergata, Via della Ricerca Scientifica, 1, 00133, Roma, Italy\label{30}\goodbreak\and
INFN Sezione di Roma2, Università di Roma Tor Vergata, via della Ricerca Scientifica, 1, 00133 Roma, Italy\label{31}\goodbreak\and
University of California, Berkeley, Department of Physics, Berkeley, CA 94720, USA\label{32}\goodbreak\and
University of California, Berkeley, Space Sciences Laboratory,  Berkeley, CA 94720, USA\label{33}\goodbreak\and
Lawrence Berkeley National Laboratory (LBNL), Computational Cosmology Center, Berkeley, CA 94720, USA\label{34}\goodbreak\and
Kavli Institute for Particle Astrophysics and Cosmology (KIPAC), Stanford University, Stanford, CA 94305, USA\label{35}\goodbreak\and
Centre Spatial de Liège (STAR Institute), University of Liège, avenue du Pré-Aily, Angleur, 4031, Belgium\label{36}\goodbreak\and
Institute of Particle and Nuclear Studies (IPNS), High Energy Accelerator Research Organization (KEK), Tsukuba, Ibaraki 305-0801, Japan\label{37}\goodbreak\and
International Center for Quantum-field Measurement Systems for Studies of the Universe and Particles (QUP), High Energy Accelerator Research Organization (KEK), Tsukuba, Ibaraki 305-0801, Japan\label{38}\goodbreak\and
Dpto. de Física Moderna, Universidad de Cantabria, Avda. los Castros s/n, E-39005 Santander, Spain\label{39}\goodbreak\and
INFN Sezione di Bologna, Viale C. Berti Pichat, 6/2 – 40127 Bologna, Italy\label{40}\goodbreak\and
Tohoku University, Graduate School of Science, Astronomical Institute, Sendai, 980-8578, Japan\label{41}\goodbreak\and
Japan Aerospace Exploration Agency (JAXA), Institute of Space and Astronautical Science (ISAS), Sagamihara, Kanagawa 252-5210, Japan\label{42}\goodbreak\and
The Graduate University for Advanced Studies (SOKENDAI), Miura District, Kanagawa 240-0115, Hayama, Japan\label{43}\goodbreak\and
Department of Physics and Astronomy, University of British Columbia, 6224 Agricultural Road, Vancouver BC, V6T1Z1, Canada\label{44}\goodbreak\and
Institut d'Astrophysique de Paris, CNRS/Sorbonne Université, Paris, France\label{45}\goodbreak\and
Lawrence Berkeley National Laboratory (LBNL), Physics Division, Berkeley, CA 94720, USA\label{46}\goodbreak\and
Laboratoire de Physique de l’École Normale Supérieure, ENS, Université PSL, CNRS, Sorbonne Université, Université de Paris, 75005 Paris, France\label{47}\goodbreak\and
Space Science Data Center, Italian Space Agency, via del Politecnico, 00133, Roma, Italy\label{48}\goodbreak\and
Université Paris-Saclay, CNRS, Institut d’Astrophysique Spatiale, 91405, Orsay, France\label{49}\goodbreak\and
Gran Sasso Science Institute (GSSI), Viale F. Crispi 7, I-67100, L’Aquila, Italy\label{50}\goodbreak\and
David A. Dunlap Department of Astronomy and Astrophysics, 50 St. George Street, Toronto ON M5S3H4, Canada\label{51}\goodbreak\and
Institute of Astrophysics, Foundation for Research and Technology-Hellas, Vasilika Vouton, GR-70013 Heraklion, Greece\label{52}\goodbreak\and
Department of Physics and ITCP, University of Crete, GR-70013, Heraklion, Greece\label{53}\goodbreak\and
INAF-Osservatorio Astronomico di Cagliari, Via della Scienza 5, 09047 Selargius, Italy\label{54}\goodbreak\and
Physics and Astronomy Dept., University College London (UCL), UK\label{55}\goodbreak\and
The University of Tokyo, Department of Astronomy, Tokyo 113-0033, Japan\label{56}\goodbreak\and
Suwa University of Science, Chino, Nagano 391-0292, Japan\label{57}\goodbreak\and
INFN Sezione di Pisa, Largo Bruno Pontecorvo 3, 56127 Pisa, Italy\label{58}\goodbreak\and
Okayama University, Department of Physics, Okayama 700-8530, Japan\label{59}\goodbreak\and
Université Paris-Saclay, CNRS/IN2P3, IJCLab, 91405 Orsay, France\label{60}\goodbreak\and
National Institute of Technology, Kagawa College, 355 Chokushi-cho, Takamatsu, Kagawa 761-8058, Japan\label{61}\goodbreak\and
Universitäts-Sternwarte, Fakultät für Physik, Ludwig-Maximilians Universität München, Scheinerstr.1, 81679 München, Germany\label{62}\goodbreak\and
NASA Goddard Space Flight Center, Greenbelt, MD 20771, USA\label{63}
}

\authorrunning{U. Fuskeland et al.}
\titlerunning{Tensor-to-scalar ratio forecasts for \\extended LiteBIRD frequency configurations}

\abstract{\LiteBIRD\ is a planned JAXA-led cosmic microwave background (CMB) $B$-mode satellite experiment aiming for launch in the late 2020s, with a primary goal of detecting the imprint of primordial inflationary gravitational waves. Its current baseline focal-plane configuration includes 15 frequency bands between 40 and 402\,GHz, fulfilling the mission requirements to detect the amplitude of gravitational waves with the total uncertainty on the tensor-to-scalar ratio, $\delta r$, down to $\delta r<0.001$. A key aspect of this performance is accurate astrophysical component separation, and the ability to remove polarized thermal dust emission is particularly important. In this paper we note that the CMB frequency spectrum falls off nearly exponentially above 300\,GHz relative to the thermal dust spectral energy distribution, and a relatively minor high frequency extension can therefore result in even lower uncertainties and better model reconstructions. Specifically, we compared the baseline design with five extended configurations, while varying the underlying dust modeling, in each of which the High-Frequency Telescope (HFT) frequency range was shifted logarithmically toward higher frequencies, with an upper cutoff ranging between 400 and 600\,GHz. In each case, we measured the tensor-to-scalar ratio $r$ uncertainty and bias using both parametric and minimum-variance component-separation algorithms. When the thermal dust sky model includes a spatially varying spectral index and temperature, we find that the statistical uncertainty on $r$ after foreground cleaning may be reduced by as much as 30--50\,\% by extending the upper limit of the frequency range from 400 to 600\,GHz, with most of the improvement already gained at 500\,GHz. We also note that a broader frequency range leads to higher residuals when fitting an incorrect dust model, but also it is easier to discriminate between models through higher $\chi^2$ sensitivity. Even in the case in which the fitting procedure does not correspond to the underlying dust model in the sky, and when the highest frequency data cannot be modeled with sufficient fidelity and must be excluded from the analysis, the uncertainty on $r$ increases by only about 5\,\% for a 500\,GHz configuration compared to the baseline.}

\maketitle

\section{Introduction}
\label{sec:introduction}

One of the key predictions of the current cosmological inflationary
paradigm is the existence of a stochastic background of primordial
gravitational waves created shortly after the Big Bang
\citep{Starobinsky:1980te,sato:1981,guth:1981,albrecht/steinhardt:1982,linde:1982,linde:1983}. 
If such waves do exist, they should induce a particular and unique
polarization signature in the cosmic microwave background (CMB) on
large angular scales, corresponding to so-called $B$-mode or
divergence-free polarization \citep{kamionskowski:1997}. Detecting this 
imprint in the CMB ranks
among the top observational priorities in modern cosmology, and huge
efforts are currently being made to develop the necessary
instrumentation, data analysis techniques, and theoretical modeling
required for this task.

Among the leading international $B$-mode efforts is \LiteBIRD\ (the Lite (Light) satellite for the study of $B$-mode polarization and Inflation from cosmic background Radiation Detection), a
satellite concept chosen by JAXA as one of their top priorities for
the coming decade, aiming for a launch from 2028 to 2029 \citep{Hazumi_2020, PTEP}. The development of
\LiteBIRD\ has already been ongoing for more than 13 years, evolving
gradually from a relatively simple and light-weight concept that
originally included only six frequency channels between 60 and
280\,GHz \citep{Hazumi:2012gjy,Matsumura_2014} into the current baseline configuration spanning 15 frequency
bands between 40 and 402\,GHz \citep{Hazumi_2020}. This increased sophistication came as a direct response to observational insights gained regarding the astrophysical sky as measured by CMB experiments such as \Planck\ \citep{PlanckI:2018} and BICEP2/Keck \citep{BICEPKeck:2016}. 
These observations pointed toward a complex reality in which both polarized synchrotron and thermal dust emission must be modeled and subtracted with exquisite precision, while simultaneously accounting for unpolarized complications such as carbon monoxide and time-dependent zodiacal light emission.

\begin{table*}[t]
  \begingroup
  \newdimen\tblskip \tblskip=5pt
  \caption{Summary of \LiteBIRD\ HFT instrument configurations considered in this paper in terms of frequency, beam size, and sensitivity. M0 denotes the baseline configuration, while M$i$ indicates a hypothetical instrument for which each HFT frequency range is scaled by a constant factor of $(1+i/10)$. }
  \label{tab:models}
  \nointerlineskip
  \vskip -3mm
  \footnotesize
  \setbox\tablebox=\vbox{
    \newdimen\digitwidth
    \setbox0=\hbox{\rm 0}
    \digitwidth=\wd0
    \catcode`*=\active
    \def*{\kern\digitwidth}
    \newdimen\signwidth
    \setbox0=\hbox{-}
    \signwidth=\wd0
    \catcode`!=\active
    \def!{\kern\signwidth}
 \halign{
   \hbox to 4cm{#\leaderfil}\tabskip 1em&
      \hfil#\hfil\tabskip 1em&
      \hfil#\hfil\tabskip 1em\hfil&
      \hfil#\hfil\tabskip 1em&
      \hfil#\hfil\tabskip 1em&
      \hfil#\hfil\tabskip 1em&
      \hfil#\hfil\tabskip 0pt\cr
    \noalign{\doubleline}
    \omit\textsc{Parameter}\hfil&
      \omit\hfil\textsc{Model}\hfil&
      \omit\hfil\textsc{Channel 1}\hfil&
      \omit\hfil\textsc{Channel 2}\hfil&
      \omit\hfil\textsc{Channel 3}\hfil&
      \omit\hfil\textsc{Channel 4}\hfil&
      \omit\hfil\textsc{Channel 5}\hfil\cr
      \noalign{\vskip 5pt\hrule\vskip 5pt}
      \noalign{\vskip 2pt}
      Frequency (GHz)     &M0& 195.0 & 235.0 & 280.0 & 337.0 & 402.0\cr
      \omit               &M1& 214.5 & 258.5 & 308.0 & 370.7 & 442.2\cr
      \omit               &M2& 234.0 & 282.0 & 336.0 & 404.4 & 482.4\cr
      \omit               &M3& 253.5 & 305.5 & 364.0 & 438.1 & 522.6\cr
      \omit               &M4& 273.0 & 329.0 & 392.0 & 471.8 & 562.8\cr
      \omit               &M5& 292.5 & 352.5 & 420.0 & 505.5 & 603.0\cr
      \noalign{\vskip 5pt\hrule\vskip 5pt}
      \noalign{\vskip 2pt}
      Beam FWHM (arcmin)  &M0& 28.6 & 24.7 & 22.5 & 20.9 & 17.9\cr
      \omit               &M1& 27.7 & 24.3 & 22.0 & 20.5 & 17.5\cr
      \omit               &M2& 26.0 & 23.1 & 21.2 & 19.9 & 17.1\cr
      \omit               &M3& 24.6 & 22.1 & 20.6 & 19.4 & 16.5\cr
      \omit               &M4& 23.5 & 21.4 & 20.1 & 18.6 & 15.6\cr
      \omit               &M5& 22.6 & 20.8 & 19.7 & 17.1 & 13.9\cr
      \noalign{\vskip 5pt\hrule\vskip 5pt}
      \noalign{\vskip 2pt}
      NET array ($\mu\mathrm{K}\sqrt{\mathrm{s}}$)  &M0& *!5.2 & *!5.3 & *!6.8 & 11 & *24\cr
      \omit               &M1& *!5.9 & *!6.4 & *!9.2 & 16 & *37\cr
      \omit               &M2& *!6.7 & *!8.0 & 12 & 23 & *60\cr
      \omit               &M3& *!7.8 & *!9.9 & 16 & 34 & 100\cr
      \omit               &M4& *!9.1 & 13 & 22 & 50 & 170\cr
      \omit               &M5& 11& 16 & 30 & 75 & 280\cr
      \noalign{\vskip 4pt\hrule\vskip 5pt} } }
  \endPlancktablewide \endgroup
\end{table*}

In the following, we revisit the upper limit of the \LiteBIRD\
frequency range, and examine its implications for the
instrument's ability to distinguish between CMB and astrophysical foregrounds. While \citet{PTEP} has already demonstrated that the baseline configuration fully achieves the primary mission goal -- namely to constrain the tensor to scalar ratio with the total uncertainty, $\delta r$, down to $\delta r < 0.001$ -- it is still desirable to optimize the uncertainties from component separation, which corresponds to one-third of the total budget error on $r$, in order to increase the overall mission margins. Exploring that is the main motivation for the current work. At the same time, we emphasize that this paper only considers the raw sensitivity and component separation aspects of the instrumental design.  We need to perform a comprehensive trade study before the
collaboration can decide on a baseline change.  The scope of the current paper is simply to determine whether sufficiently significant gains are available to warrant such a study. 
 
To this end, we compared the current baseline configuration with five
alternative configurations in which the upper limit was varied between
400 and 600\,GHz in steps of 40\,GHz. The absolute upper center
frequency limit of 600\,GHz was dictated by the current bolometer
design, which has a hard cutoff at 680\,GHz \citep{novotny1975single}. 
To additionally
minimize the number of changes required for the overall satellite
design, we chose to only modify one of the three \LiteBIRD\ telescopes,
namely the High-Frequency Telescope (HFT); the Low- and Mid-Frequency Telescopes 
(LFT and MFT) have been left unchanged. 
Frequencies up to 600\,GHz can be considered only because \LiteBIRD\ is a space mission, while ground-based experiments are restricted to frequencies below 300\,GHz due to the atmospheric windows \citep{liebe1981,pardo2001}.

The main question considered in this paper is the quantitative
relationship between the frequency range and the tensor-to-scalar ratio
uncertainty marginalized over polarized thermal dust
emission. Previous analyses \citep[e.g.,][]{remazeilles:2016} have
suggested that a low cutoff on the high frequency side only supports very limited
ability to constrain the spectral energy distribution (SED) of thermal dust
emission. At the same time, it is well known that the CMB SED falls
nearly exponentially above 300\,GHz relative to the thermal dust SED
\citep[see e.g., figure 35 of][]{Planck2018compsep}, and therefore even
relatively small changes in the cutoff frequency may have a dramatic
impact on the confusion between CMB and thermal dust.  Quantifying the
importance of including frequencies higher than where the CMB SED
drops off in a realistic setting is the main goal of the current
paper. We note that while we subsequently address \LiteBIRD\ in
particular, the main arguments are instrument-agnostic, and the
primary conclusions are therefore generally applicable to any future
space mission or balloon-borne experiment. Indeed, similar analyses have recently been published for both the PICO \citep{aurlien:2022} and ECHO \citep{sen:2022} CMB satellite concepts with consistent results.

This challenge of optimizing the frequency range may be separated into
two important and complementary aspects, namely optimizing our
sensitivity to the tensor-to-scalar ratio overall and maximizing our ability
to reject wrong sky models. The following analysis is organized
accordingly. To address the first question, we analyze ideal
foreground simulations in which the fitting model matches the
simulated sky, and estimate the tensor-to-scalar ratio uncertainty as
a function of frequency coverage. In this case, we adopt a standard
one-component modified blackbody (MBB) thermal dust model with
spatially varying spectral parameters, following closely in the
footsteps of \citet{PTEP}. In order to ensure robustness in terms of
analysis-dependent details, we employed five independent component-separation
methods for this task, three of which implement different
variations of parametric fitting (\commander\ -- \citealp{eriksen:2008},
\FGBuster\ -- \citealp{Errard2018}, 
and Moment Expansion -- \citealp{Vacher2022}), 
while the last two implement
(semi)blind internal linear combination (ILC) fitting (Needlet ILC, \NILC\ and constrained moment ILC, \cMILC -- \citealp{Remazeilles2011,Remazeilles2021}).

To assess the capability of rejecting incorrect data models, we
analyzed simulations based on an extended thermal dust model, namely
the physically motivated silicate-iron-carbon model (SiFeC;
\citealp{hensley:2017}). We evaluated the residual
$\chi^2$, and compared this with the derived tensor-to-scalar
estimates. For this study, we note that only \commander\ currently
provides $\chi^2$-based goodness-of-fit estimates, and we therefore
only reported results for that code in this part of the paper. However, we note
that work is currently ongoing on implementing similar statistics for
other methods, and those results will be reported elsewhere.

It is also important to note that the foreground sky in this paper was approximated as a ``single layer model,'' and no attempt was made to take into account the full 3D structure of the Milky Way. Full line-of-sight integration of 3D effects would introduce additional frequency-decorrelation effects \citep{tassis2015, Chluba2017,pelgrims2021,ritacco2022,Vacher2022c}, and the following results therefore represent an optimistic view of the true sky in terms of complexity. In reality, there will be additional structures, both spatially and spectrally, and fully resolving these would in principle require even more data than considered here, for instance in terms of additional frequency bands, wider frequency range, or dedicated 3D constraints on the Galactic magnetic field. The following optimization results should therefore be considered as a lower bound on what is actually needed for a future production analysis.

The paper is organized as follows. In Sect.~\ref{sec:litebird} we
provide a brief overview of the \LiteBIRD\ concept and current
baseline, and we define a set of alternative configurations to be
considered. In Sect.~\ref{sec:simulations} we describe the simulations
used for the present analysis, and in Sect.~\ref{sec:algorithms} we
briefly survey the component-separation methods used in this paper;
algorithmic details are deferred to the Appendices. In Sect.~\ref{sec:mbb}, we
present the results derived from an ideal model analysis, while in
Sect.~\ref{sec:complex_dust} we consider modeling errors. 
Final conclusions are drawn in 
Sect.~\ref{sec:conclusions}.

\section{Baseline and extended LiteBIRD instrument configurations}
\label{sec:litebird}

In this section, we discuss the \LiteBIRD\ mission characteristics
that are relevant to the present analysis. A comprehensive review of
the instrument concept can be found in \cite{PTEP}.  First, the
\LiteBIRD\ detector arrays are populated with a total of 4508 lenslet-
or horn-coupled transition edge sensor (TES) bolometers fabricated on 
silicon wafers, covering
15 frequency bands between 40 and 402\,GHz.  The \LiteBIRD\ bolometers
provide exquisite sensitivity across this full frequency range, but
the technology itself is limited to frequencies <\,678\,GHz,
corresponding to the superconducting gap of niobium -- the material
used for mm-wave transmission lines on the detectors
\citep{novotny1975single}.  This value therefore naturally defines a
clear upper limit for the feasible frequency range of the
\LiteBIRD\ instrument in any modification to the baseline.

\begin{figure}
  \begin{center}
    \includegraphics[width=\linewidth]{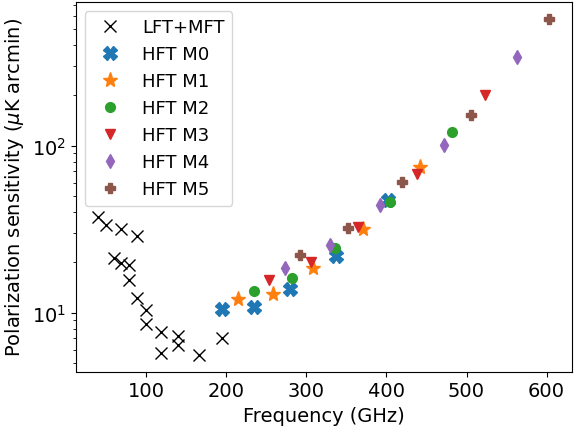}
  \end{center}
  \caption{Sensitivity versus frequency for the various \litebird\ configurations listed in Table~\ref{tab:models}. The black crosses indicate LFT or MFT frequencies that are not modified in this paper, while colored symbols indicate the various modified HFT frequencies.}
  \label{fig:hfsl_sens}
\end{figure}

As a second consideration, to ensure robust separation between the actual 
polarization signal and a wide range of intensity-specific contaminants,
\LiteBIRD\ will employ half-wave plates (HWPs) spinning at a rate of $0.5$ to $0.8$\,Hz.
Important strengths of mesh filter-based HWPs are low
mass requirements and high transmission. However, an important
drawback is finite effective bandwidth, typically corresponding to a
ratio between its maximum and minimum effective frequency of about 2.3
\citep{PTEP}. As such, the full relative \LiteBIRD\ frequency range of
402\,GHz/40\,GHz$\,\approx\,$10 cannot be supported within a single
telescope, but rather three separate telescopes are required. For
\LiteBIRD\, these are called the Low-, Mid-, and High-Frequency
Telescopes, respectively, or LFT, MFT, and HFT as a short-hand. While
this organization comes at a significant cost in terms of system
complexity, it also allows significant optimization of each system,
and different technologies may be used as appropriate.

In total, the three telescopes span 22 independent frequency channels,
covering 15 frequency bands; there is an overlap of three channels
inside the LFT, three channels between the LFT and MFT, and an overlap
of one channel between the MFT and HFT. These overlaps represent
important cross-checks against systematic errors in any of the three
telescopes. Each frequency band overlaps the adjacent bands by
approximately half the bandwidth. The radiation is detected by seven
different wafer types, each containing one or two types of
multichromatic pixels. Each pixel, for each polarization state,
distributes power through one, two, or three different bandpass
filters to independent TES detectors. The HFT detectors are
implemented in terms of three different detector modules; one with
3-color pixels, one with 2-color pixels, and the highest frequency
band on single-color pixels \citep{Montier_2020}.

The main goal of the current paper is to investigate the optimal 
\LiteBIRD\ frequency range with respect to the removal of the thermal 
dust from the CMB polarization signal.
Given the mature state of the current baseline, 
modifications must be realistic and the scientific improvements for doing this must be substantial.
One of the parameters that may still be adjusted, however, is the
effective HFT frequency range. Specifically, it is possible to shift
the entire frequency range by a constant factor, which ensures that
the relative factor between the highest and lowest frequency bands is
unchanged, as required by the mesh-filter HWP. In this paper, we
therefore defined a set of extended HFT configurations parametrized by
a scaling factor, $f$, that varied between 1.0 and 1.5 in steps of
0.1, and shifted the entire HFT frequency range. These models were
denoted M$i$, such that M0 corresponded to the baseline, while M5
corresponded to an extended model with a highest frequency of
$1.5\times 402\,\mathrm{GHz} \approx 603\,\mathrm{GHz}$. For each
modified frequency channel, we calculated new instrumental parameters,
such as sensitivity or beam size, using the same methodology as used
for the baseline in \cite{PTEP}. This included evaluating the system
temperature through ray tracing from the feed to terminals, either
inside the telescope or the sky. The beam full width at half maximum (FWHM) values were estimated
via the edge taper at the telescope aperture, as calculated from the
feed far-field pattern. The resulting values are summarized
numerically in Table~\ref{tab:models}, while Fig.~\ref{fig:hfsl_sens}
shows the sensitivity as a function of frequency for the extended
bands.

It is interesting to note that the sensitivity of two channels in
different configurations can vary significantly, despite the fact that
both their center frequencies and detector counts are nearly
identical; a typical example of this is channel 1 in M2 and channel 2
in M0. The reason for this is that the optics for the HFT are optimized for the telescope’s full frequency bandwidth, not any single channel.  Frequencies near the edge of the telescope bandwidth typically have poorer optical absorption and reflection properties than frequencies nearer the center of the band.

We also note that the LFT and MFT are left unchanged in this exercise, and
their instrumental parameters are not listed in
Table~\ref{tab:models}. The primary motivation for this choice is
programmatic; changing the baseline at this late stage will carry a
nonnegligible cost, and it is therefore highly desirable to make as
few changes as possible. At the same time, it is important to note
that this choice does come at a cost in the form of weaker
systematics cross-check between the MFT and HFT, since the two telescopes
no longer have overlapping frequency ranges. Thus, these modified
configurations may be considered as trading off robustness toward
``unknown unknowns'' (for instance, subtle HWP frequency effects)
against robustness to a known unknown (polarized thermal dust
emission). If this issue is considered sufficiently important at a
programmatic level, it is of course also possible to modify the
lower-frequency telescopes, albeit at a somewhat higher cost.

\section{Simulations}
\label{sec:simulations}

\begin{figure}
  \begin{center}
    \includegraphics[width=\linewidth]{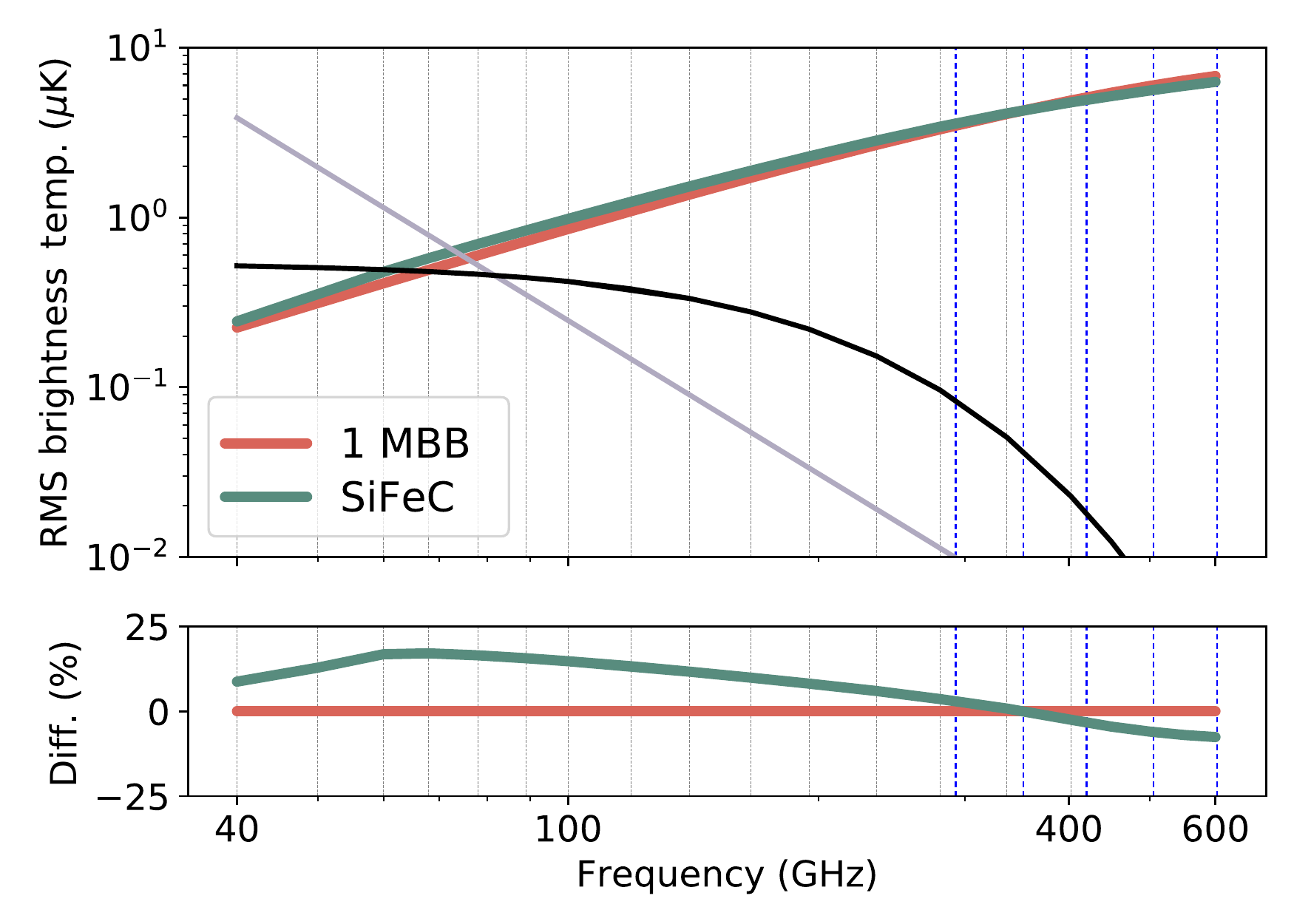}
  \end{center}
  \vspace{-.3cm}
  \caption{SEDs of the two dust models considered in this work, the 1
    MBB, (red line) and the physical dust model SiFeC, (green line). 
    The black line
    shows the SED of the CMB, while the gray refers to synchrotron
    emission. All the SEDs are in brightness temperature units and are
    computed considering the rms of Stokes $Q$ and $U$ maps on 70\,\% of 
    the sky. Sky masks are obtained by retaining the cleanest 
    fraction of the sky in polarized intensity at 100 GHz. Vertical
    gray lines show the central frequency of the baseline
    \LiteBIRD\ frequency channels, and in blue dashed lines we report the
    central frequency of the HFT M5 extension. The bottom panel shows
    the percentage difference of SiFeC with respect to MBB. }
  \label{fig:dust_sed}
\end{figure}

The \LiteBIRD\ simulations considered in this paper are generated
directly in pixel space, and account only for sky signal and white
noise. The sky emission components for all 22 frequency channels are
generated using the PySM (Python Sky Model; \citealp{pysm})
package. The LFT and MFT channels are defined according to the
baseline instrument configuration in \cite{PTEP}, while the HFT
channels are generated in six versions, called M0--M5, as tabulated in
Table~\ref{tab:models}. Here, M0 is the \LiteBIRD\ baseline, and M1--M5
are the frequency extended data sets. The instrumental summary
parameters listed in this table are derived from first principles
using the same methodology as the baseline configuration in
\citet{PTEP}, properly accounting for realistic optics, bandpass, and
bolometer effects.

We simulate the polarized microwave emission considering three
different components, namely a cosmological CMB signal and both
diffuse synchrotron and thermal dust radiation from our own
Galaxy. The CMB component is simulated as a Gaussian random
realization of the \Planck\ 2018 $\Lambda\textrm{CDM}$ best-fit power
spectrum \citep{planck2018param}. In the following, we set the 
tensor-to-scalar ratio, $r$, to zero. Obviously, a main goal of 
\LiteBIRD\ is to actually measure a non-zero value of $r$, and it 
could therefore be of interest to also consider non-zero values of 
this parameter. However, this paper is primarily concerned with the 
relative performance of different instrument configurations. This 
comparison is not sensitive to the exact value of $r$, and then for 
the purpose of our analyses we decided to use $r=0$ as reference, 
which also provides the benefit of not having to deal with the 
cosmic variance noise term. Future work focusing on absolute 
detection level performance should obviously revisit this with a 
range of alternative cosmological parameter values, including $r$, 
the optical depth of reionization $\tau$, and others.

Galactic synchrotron radiation is physically generated by cosmic-ray
electrons that are accelerated by the Galactic magnetic field. This
signal is highly polarized with a polarization fraction that can reach
20\,\% at intermediate and high Galactic latitudes
\citep{Planck2016lowfreq,Kogut2007WMAP3}. Synchrotron emission
represents the dominant foreground at frequencies below about 70\,GHz
\citep{Krachmalnicoff2016, Planck2016compsep}. As a first
approximation, the synchrotron SED
follows a power law with a spatially varying spectral index 
\citep{lawson:1987, platania:1998, bennett:2003}.
Recent analyses support a spectral index of $\beta_\mathrm{s}\approx-3$
\citep{Fuskeland2014, Fuskeland2021}, 
with variations on the order of 10\,\% across the sky on degree 
angular scales \citep{Krachmalnicoff2018}. 
In the following, we adopt the \rm{s}1
PySM model for synchrotron emission, corresponding to a strict
power-law spectral behavior with spatially varying spectral index. We
note that synchrotron emission is largely irrelevant for the highest
\LiteBIRD\ frequencies, and any conclusions regarding the optimal HFT
frequency range will be largely independent of the synchrotron model.

The same does, however, not hold true for thermal dust emission, which
is the dominant foreground component at frequencies above 70\,GHz
\citep{Krachmalnicoff2016, Planck2016compsep}. Thermal dust emission
is generated by vibrating nonspherical dust grains in the
interstellar medium (ISM), 
and is polarized because the small axes of the grains are preferentially aligned
parallel to the local magnetic field \citep{hoang2016,hensley2022astrodust}. 
The detailed physical processes involved in
this are, however, more complicated than for synchrotron emission, and
the predicted emission from any region of the ISM will depend on a
wide range of local properties, including dust grain composition, the 
local radiation field and magnetic field structure. For robustness,
we therefore consider two different thermal dust SED models in the
following, namely; (1) a phenomenologically motivated single modified
blackbody parametrized by a spatially varying spectral index and
temperature (1 MBB; PySM model \rm{d}1); and (2) a physically motivated
silicate-iron-carbon model (SiFeC; PySM model \rm{d}8,
\citealp{hensley:2017}).  The 1 MBB model is very often
used in the literature. The second model is more exotic, and is
physically motivated rather than data-driven. They also differ in
complexity in terms of spatial variation. The \rm{d}1 model has both a
spatially varying spectral index and temperature,
while \rm{d}8 has spatially constant spectral parameters. 
Future work should also consider models that include physically motivated 
frequency decorrelation, for instance by implementing proper multi-layer 
3D models. This is however beyond the scope of the current paper. 

Figure~\ref{fig:dust_sed} shows the dust SEDs for the two different
models computed on $70$\,\% of the sky. Sky masks are obtained by 
retaining the cleanest fraction of the sky in polarized intensity at 
100\,GHz. As seen in the bottom panel,
the relative difference between the two models can reach tens of percent
across the \LiteBIRD\ frequency range, and properly accounting for these
variations will be essential for making a robust CMB extraction. For
comparison, the gray and black curves show the synchrotron and CMB
SEDs. The near exponential drop-off of the CMB spectrum is clearly
seen above about 300\,GHz, leading to a rapidly increasing ratio
between thermal dust and CMB emission between 400 and 600\,GHz.

Instrumental effects are modeled under highly idealized assumptions,
since these
are considered to be subdominant regarding the central
question of frequency range versus thermal dust reconstruction, and
rather only add computational complexity and cost. In particular, we
model all instrumental bandpasses in terms of Dirac $\delta$ functions, all
instrumental beams as azimuthally symmetric Gaussians, and the noise
is assumed to be uncorrelated, Gaussian, and spatially
isotropic. Center frequencies, beam FWHMs, and array noise equivalent
temperatures (NETs) are all listed in Table~\ref{tab:models}. Under these
assumptions, the simulation procedure is very straightforward, and does
not require any low-level time-domain processing. 

The data model for a given frequency band $\nu$ may be described as 
\begin{linenomath*} 
\begin{equation} 
    \vec{d}_{\nu} = \vec{s}_{\nu} + \vec{n}_{\nu}, \label{eq:data_model}
\end{equation}
\end{linenomath*} 
where $\vec{d}$ represents a given (simulated) sky map, $\vec{s}$ denotes the sky signal, and $\vec{n}$ is instrumental noise. 
The simulated sky maps analyzed in this paper are based on two different thermal dust emission models. For the baseline model, we only fit the one-component MBB model, while the SiFeC model is used to assess modeling errors and is therefore only present in the simulations, and not in the data model in the methods. Including also CMB and synchrotron emission, Eq.~(\ref{eq:data_model}) may then be written in the following explicit form, adopting thermodynamic CMB temperature units,
\begin{linenomath*} 
\begin{align} 
    \vec{d}_{\nu} & = \vec{a}_{\mathrm{CMB}} \\
    & + \vec{a}_{\mathrm{s}}(\nu)\ \gamma(\nu ) \left( \frac{\nu}{\nu_{0,\mathrm{s}}} \right)^{\beta_\mathrm{s}} \\
    & + \vec{a}_{\mathrm{d}}(\nu )\ \gamma(\nu ) \frac{e^{\frac{h\nu_{0,\mathrm{d}}}{kT_{\mathrm{d}}}}-1}{e^{\frac{h\nu}{kT_{\mathrm{d}}}-1}} \left(\frac{\nu}{\nu_{0,\mathrm{d}}} \right)^{\beta_{\mathrm{d}}+1} \label{eq:dust_model_1}\\
    & + \vec{n}_{\nu},
    \label{eq:data_model_expanded}
\end{align}
\end{linenomath*} 
where $\{\vec{a}_{\mathrm{CMB}},\vec{a}_{\mathrm{s}}, \vec{a}_{\mathrm{d}}\}$ are the signal amplitudes relative to the reference frequency $\nu_{0}$ for each signal, $\gamma(\nu)$ is the conversion factor between the Rayleigh-Jeans brightness temperature and the CMB thermodynamic temperature, $\{\beta_{\mathrm{s}}, \beta_{\mathrm{d}}\}$ are the spectral indices, $T_{\mathrm{d}}$ is the temperature for thermal dust, and $h$ and $k$ are the Planck and Boltzmann constants, respectively. In principle, this sky model is valid for both intensity and polarization, with individual parameters fitted for each Stokes parameter. In practice, however, we only include CMB, synchrotron and thermal dust emission in polarization, and we additionally assume $\beta_Q = \beta_U$.

All simulations are generated both at full angular resolution
(adopting a {\tt HEALPix} pixelization with a resolution parameter of
$N_{\mathrm{side}}=512$; \citealp{healpix}) and at low angular
resolution $(10^{\circ}$ FWHM, $N_{\mathrm{side}}=16$). The former set is used by the Moment
expansion, \FGBuster\ and \NILC/\cMILC\ analyses, while the latter is
used by the \commander\ analysis. We note that the low-resolution maps
do not include subpixel (or subbeam) structure, but are generated
natively at the target resolution. The motivation for this choice is
that sub-pixel structure may excite spurious $B$-mode power, both from
beam averaging and from parallel transport inaccuracies during {\tt HEALPix}
downgrading. At the same time, a future \LiteBIRD\ \commander\ analysis
will be performed in the time-domain with full angular resolution data,
following a methodology similar to that described by \citet{beyondplanck2020},
and this complication will therefore not be relevant for the final
\LiteBIRD\ analysis.

\section{Component-separation algorithms}
\label{sec:algorithms}

In order to ensure that the general results are robust with respect to
algorithmic details, we employ a total of five different component-separation
algorithms in the following. These may be divided into two
main groups, namely parametric (\commander, \FGBuster, and Moment
Expansion) and (semi)blind methods (\NILC\ and \cMILC).

\begin{figure}
  \begin{center}
    \includegraphics[width=\linewidth]{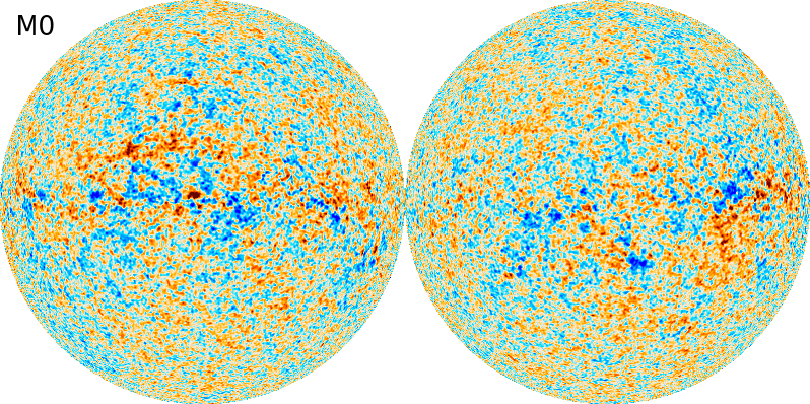}
    \includegraphics[width=\linewidth]{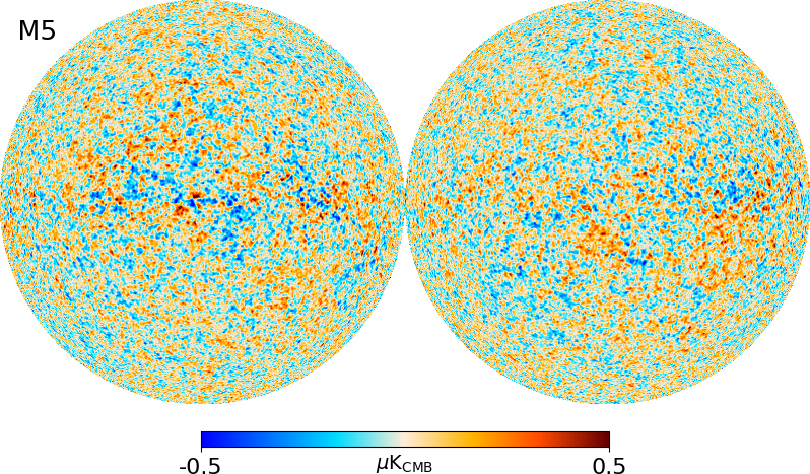}
  \end{center}
  \caption{Comparison of the reconstructed CMB $B$-mode maps from
    \nilc\ for the M0 (\emph{top row}) and M5 (\emph{bottom row})
    instrument configurations. The two panels in each row show opposite
    hemispheres aligned with the Galactic plane along the equator, and
    the north and south Galactic poles at the top and bottom,
    respectively. The left and right panels are centered on the
    Galactic center and anticenter.}
  \label{fig:nilc_map}
\end{figure}

\begin{figure*}
  \begin{center}
    \includegraphics[width=0.235\linewidth]{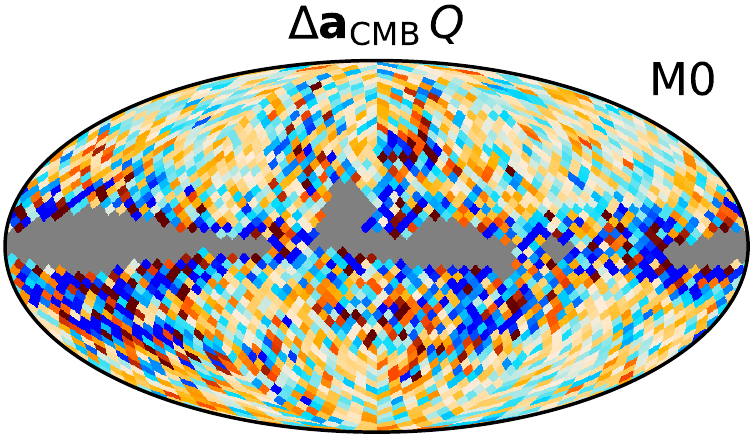}
    \includegraphics[width=0.235\linewidth]{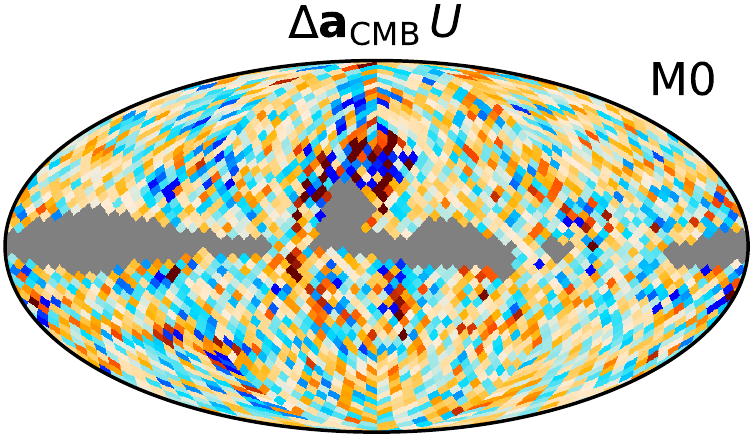}\hspace*{5mm}
    \includegraphics[width=0.235\linewidth]{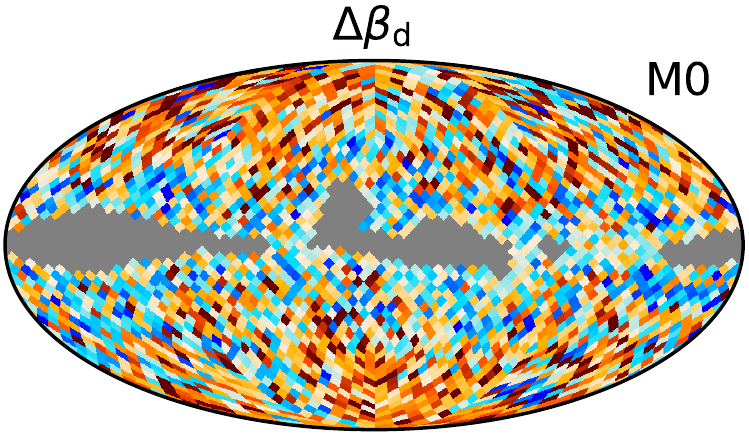}
    \includegraphics[width=0.235\linewidth]{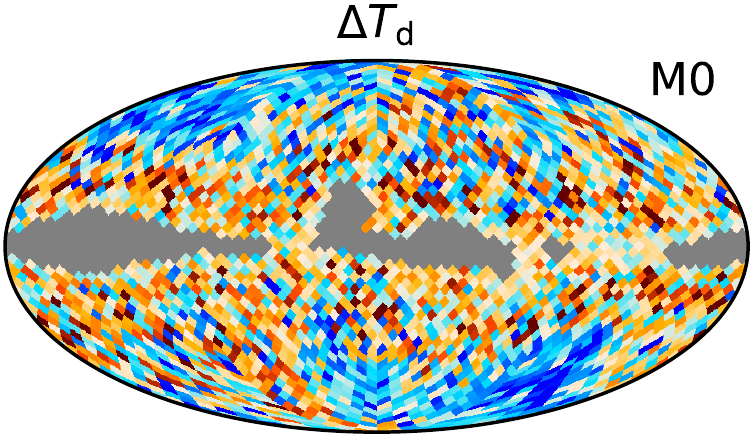}\\
    \includegraphics[width=0.235\linewidth]{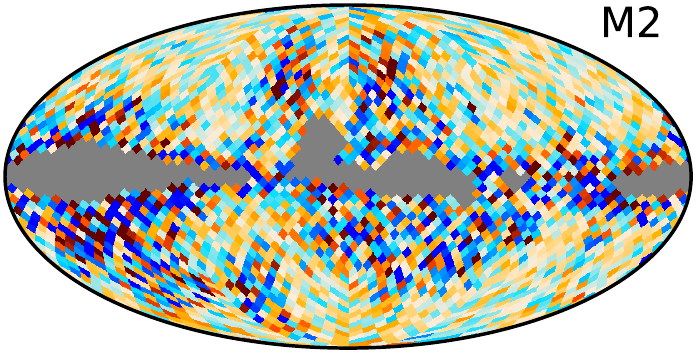}
    \includegraphics[width=0.235\linewidth]{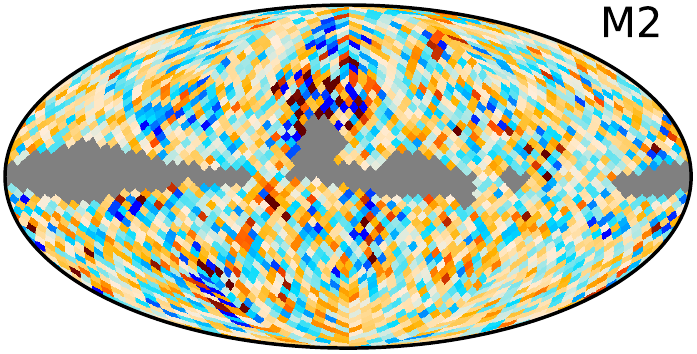}\hspace*{5mm}
    \includegraphics[width=0.235\linewidth]{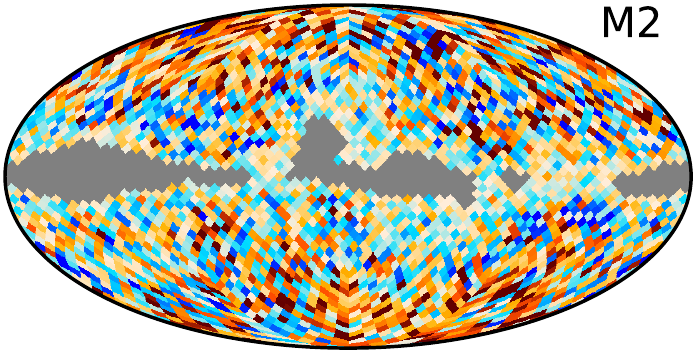}
    \includegraphics[width=0.235\linewidth]{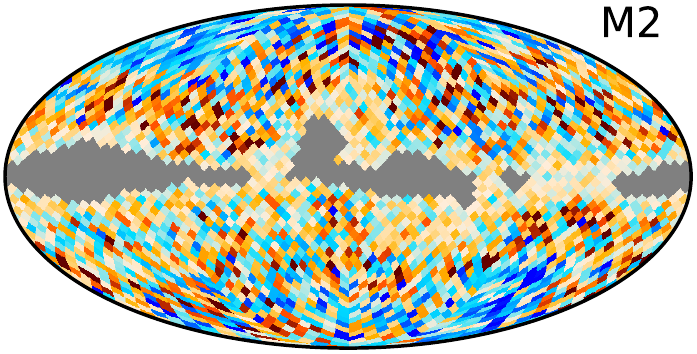}\\
    \includegraphics[width=0.235\linewidth]{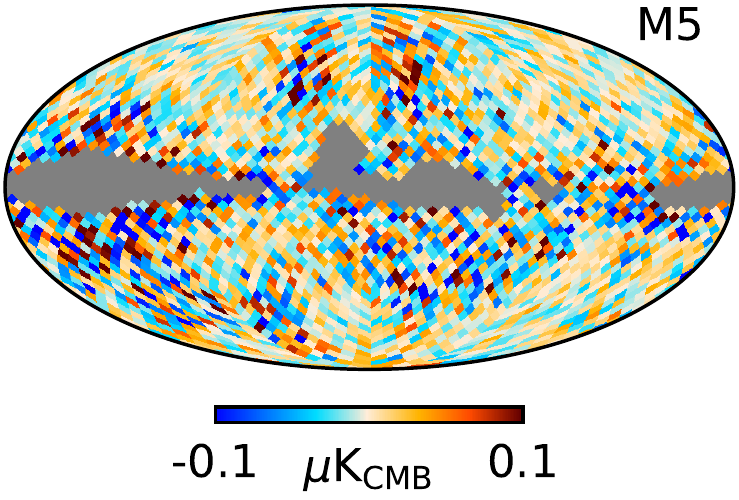}
    \includegraphics[width=0.235\linewidth]{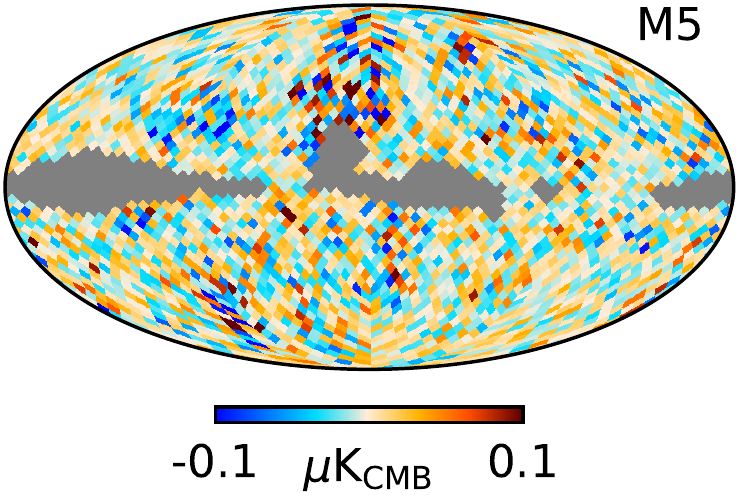}\hspace*{5mm}
    \includegraphics[width=0.235\linewidth]{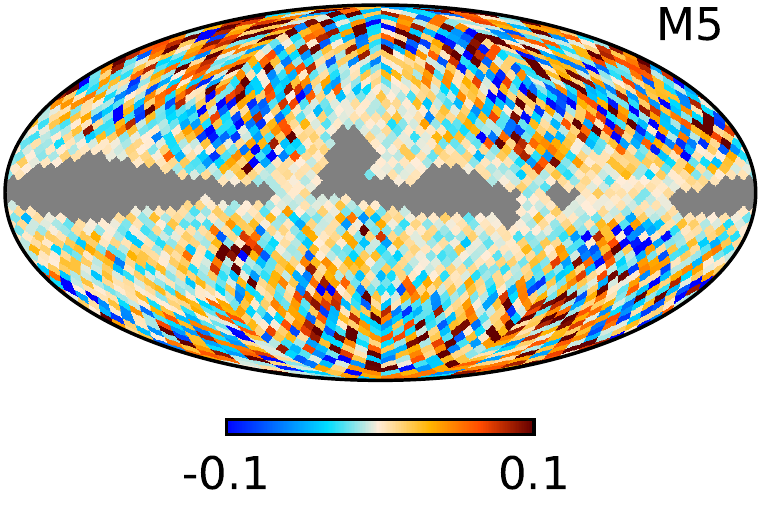}
    \includegraphics[width=0.235\linewidth]{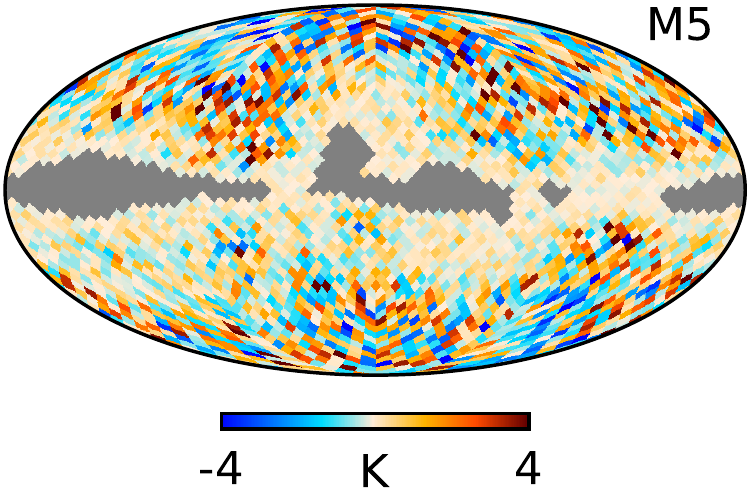}
  \end{center}
  \caption{\commander\ residual (output minus input) maps for selected parameters and instrument configurations for one arbitrarily selected Monte Carlo sample. Columns show, from left to right; (1) CMB Stokes $Q$; (2) CMB Stokes $U$; (3) thermal dust spectral index; and (4) thermal dust temperature. Rows show M0, M2, and M5. The maps show one typical realization. Gray pixels indicate the masked pixels with $f_{\mathrm{sky}}=90\,\%$.}
  \label{fig:comm_param_residuals}
\end{figure*}

\Commander\ \citep{eriksen:2008} is a Bayesian Gibbs sampler that has
been successfully applied to \Planck, \WMAP, \LiteBIRD, and many other
data sets \citep[e.g.,][]{Planck2016compsep}. The defining feature of
this method is an explicit joint parametric model that simultaneously
accounts for cosmological and astrophysical parameters, and this global
model is explored by textbook Markov chain Monte Carlo methods. A
major strength of this method is the availability of well-defined
goodness-of-fit $\chi^2$ statistics and realization-specific uncertainty
estimates, while a major weakness is a high computational cost. In
this paper we only apply this method to low-resolution simulations;
for further details, see Appendix~\ref{app:appendix_comm}.

\FGBuster\ \citep{Puglisi2022} is based on a
similar statistical foundation as \Commander, but uses nonlinear
optimization to explore the likelihood function rather than Monte
Carlo sampling. It further speeds up the estimation process by first
estimating all nonlinear spectral foreground parameters marginally
with respect to linear amplitude parameters; then it estimates
amplitudes conditionally with respect to the spectral
parameters. Finally, angular power spectra are computed from the
amplitude maps. This method has been used extensively for \LiteBIRD\
forecasting and optimization \citep[e.g.,][]{PTEP}. For further
details, see Appendix~\ref{app:appendix_fgbuster}.

The moment-expansion method \citep{Chluba2017,Vacher2022b} also has a similar
statistical foundation as \Commander\ and \FGBuster\, but it allows for
Taylor-expansion-based distortions in the SED models for 
each foreground to account for nonlinear averaging
effects, for instance from line-of-sight integration, beam convolution or harmonic expansion. This method has already been applied to baseline \LiteBIRD\
simulations by \citet{Vacher2022}. For further details, see
Appendix~\ref{app:appendix_mom}.

While each of the above methods is based on explicit and nonlinear
parametric foreground models, the Needlet Internal Linear Combination
(\nilc; \citealp{Delabrouille2009}) method takes a fundamentally
different approach, and assumes simply that the total foreground
signal may be written as a linear combination of the true CMB signal
and the total foreground signal at each frequency channel, and then
form the linear combination of frequency maps that minimizes the
variance of the final product, see Appendix~\ref{subsec:ilc}. 
To account for possible spatial
variations in the foreground SED, the linear combination weights are
computed separately in needlet space, allowing for localized
optimization both in terms of sky position and angular scale. This
method has been applied to a wide range of data sets, including
\Planck\ \citep[e.g.,][]{Planck2016compsep} and \LiteBIRD. For further
details, see Appendix~\ref{subsec:nilc}.

Finally, we consider the more recently developed constrained moments
ILC method (\cmilc; \citealp{Remazeilles2021}), which combines the
moment-expansion and \NILC\ methods. Specifically, additional
constraints are imposed on the ILC weights that explicitly cancels out
individual foreground components through their Taylor-expanded
SEDs. Each moment corresponds to one additional linear constraint in
the ILC solution, and the combination of foreground-specific and the
variance constraints is solved through a single Lagrange multiplier
system. For further information, see Appendix~\ref{subsec:cmilc}.

\section{Ideal model analysis: Estimation of statistical uncertainties}
\label{sec:mbb}

When contemplating a major modification of a satellite's baseline design,
there are at least two aspects that must be considered
carefully. The first aspect is sensitivity -- 
we must investigate how much stronger (or weaker) constraints on the 
tensor-to-scalar ratio the proposed modification will lead to.
The second aspect is goodness-of-fit -- 
we must also investigate whether the proposed modification will affect our 
sensitivity to modeling errors or our ability to detect such errors.
These two main aspects are addressed individually in this and the next section.

\begin{figure*}
  \begin{center}
    \includegraphics[width=0.49\linewidth]{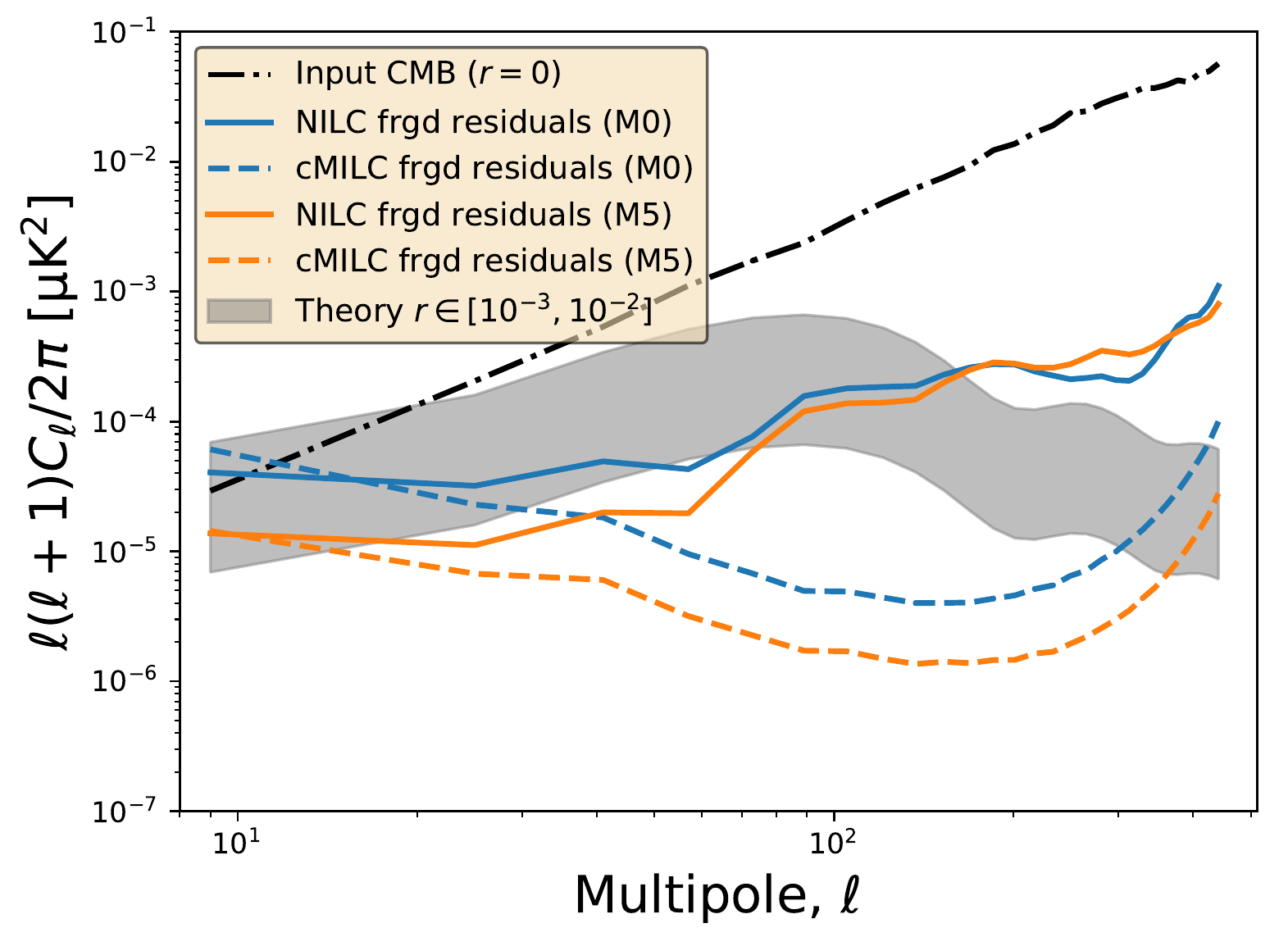}
    \includegraphics[width=0.49\linewidth]{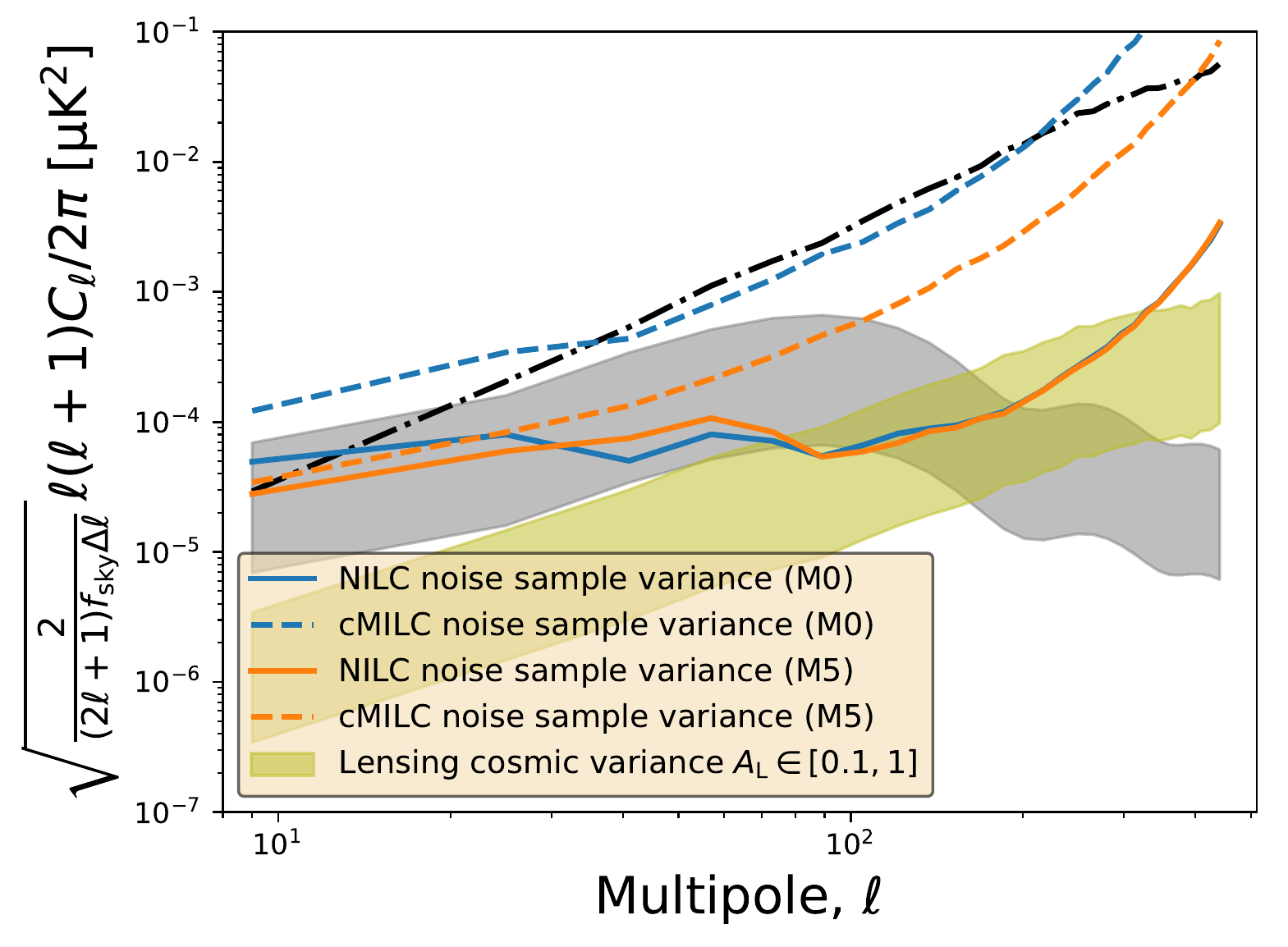}
  \end{center}
  \caption{Power spectrum of residual foreground contamination
    (\emph{left panel}) and noise (\emph{right panel}) with
    \nilc\ (solid lines) and \cmilc\ (dashed lines) on $f_{\rm
      sky}=50\,\%$ of the sky for the M0 (blue) and M5 (orange)
    configurations. For reference, we show the input CMB $B$-mode
    signal ($r=0$, black dash-dotted line), the primordial $B$-mode
    signal expected from theory for values ranging from $r=10^{-2}$
    down to $r=10^{-3}$ (gray-shaded area), and  the residual lensing
    cosmic variance (yellow-shaded area) for no delensing ($A_\mathrm{L}=1$) up
    to 90\,\% delensing ($A_\mathrm{L}=0.1$). All the spectra are binned with a multipole bin size of $\Delta\ell=16$.
  }
  \label{fig:nilc_ps}  
\end{figure*}

Starting with the sensitivity aspect, we approach this by analyzing
the simulations summarized in Sect.~\ref{sec:simulations} for each of
the instrument configurations described in
Sect.~\ref{sec:litebird}. In this section, we primarily focus on the
simplest foreground model, namely the one-component MBB thermal dust
model, for which modeling errors are minor (it is worth noting that
they are by no means nonexistent, since both informative priors and
spatial variations in the fitted parameters can lead to biases). 

We also note that each of the five component-separation methods
discussed in Sect.~\ref{sec:algorithms} have their particular
strengths and weaknesses, and our goal in this paper is not to perform
a head-to-head component-separation algorithm comparison, but it is
rather to assess the capabilities and limitations of the \LiteBIRD\
frequency selection itself. In the following, we therefore present
selected results from among the five methods, depending on which
product is most convenient for a particular application. In general,
however, we note that all five methods provide qualitatively similar
results.

\subsection{Map residuals}

First, to build intuition regarding the overall impact of the
frequency range, we show in Fig.~\ref{fig:nilc_map} the reconstructed
$B$-mode \nilc\footnote{Note that \nilc, unlike the parametric
  methods, is defined natively in $EB$ space rather than Stokes $QU$
  space.} map for one arbitrarily selected simulation and each of the
two most extreme instrument configurations, M0 (top row) and M5
(bottom row). While residual foreground contamination is clearly
visible in the Galactic disk region of the CMB $B$-mode map for M0, we
see that these residuals are significantly reduced for the M5
configuration, with a wider frequency range.

A complementary view from the parametric \Commander\ code is
shown in Fig.~\ref{fig:comm_param_residuals} in the form of residual
(output-minus-input) maps for four key quantities, namely the Stokes
$Q$ and $U$ CMB maps and the thermal dust spectral index and
temperature for the M0, M2, and M5 instrument
configurations.\footnote{For visualization purposes, the gray pixels
  in this figure indicate a 90\,\% analysis mask, which is
  significantly smaller than the default 73\,\% confidence mask
  adopted by \Commander\ for the likelihood analysis.} In all cases,
we see that the quality of the fit improves with increasing maximum
frequency. In the CMB maps, this is most clearly seen in the form of
reduced scatter around the Galactic plane, while for $T_{\mathrm{d}}$
the dark blue area in the high Galactic latitudes systematically fade
from M0 to M5; correspondingly, the dark red area in
$\beta_{\mathrm{d}}$ also gradually fade away. These observations may
be understood intuitively by noting that an increased frequency range
improves the ability to break the well-known degeneracy between
$\beta_{\mathrm{d}}$ and $T_{\mathrm{d}}$ \citep[e.g.,][]{juvela_ysard_2012}
that arises when fitting for them jointly.

\subsection{Power spectrum uncertainties}

Next, we consider the impact of changing the instrument configuration
from M0 to M5 in terms of residual foreground and noise angular power
spectra. This is shown in Fig.~\ref{fig:nilc_ps} for both
\nilc\ (solid lines) and \cmilc\ (dashed lines). The left panel shows
the projected foreground power spectrum, while the right panel shows
the noise sample variance. First, we see that \cmilc\ provides a
significantly lower residual foreground contamination than
\nilc\ across a wide range of multipoles for both instrument
configurations. This is due to the additional deprojection of
foreground moments, which is more efficient at suppressing true
foreground residuals. However, this additional foreground reduction
performance does come at a significant price in the form of higher
noise uncertainties caused by the same additional degrees of
freedom. This noise penalty is greatly reduced when extending the
\LiteBIRD\ frequency range from 402 to 603\,GHz, to the point that it
is comparable to the CMB cosmic variance. In addition, the overall
level of residual foreground contamination is also greatly reduced for
the extended frequency range, both for \nilc\ and \cmilc, across all
multipoles.

\subsection{Tensor-to-scalar ratio constraints}
\label{sec:r_constraints}

\begin{figure}
  \begin{center}
    \includegraphics[width=\linewidth]{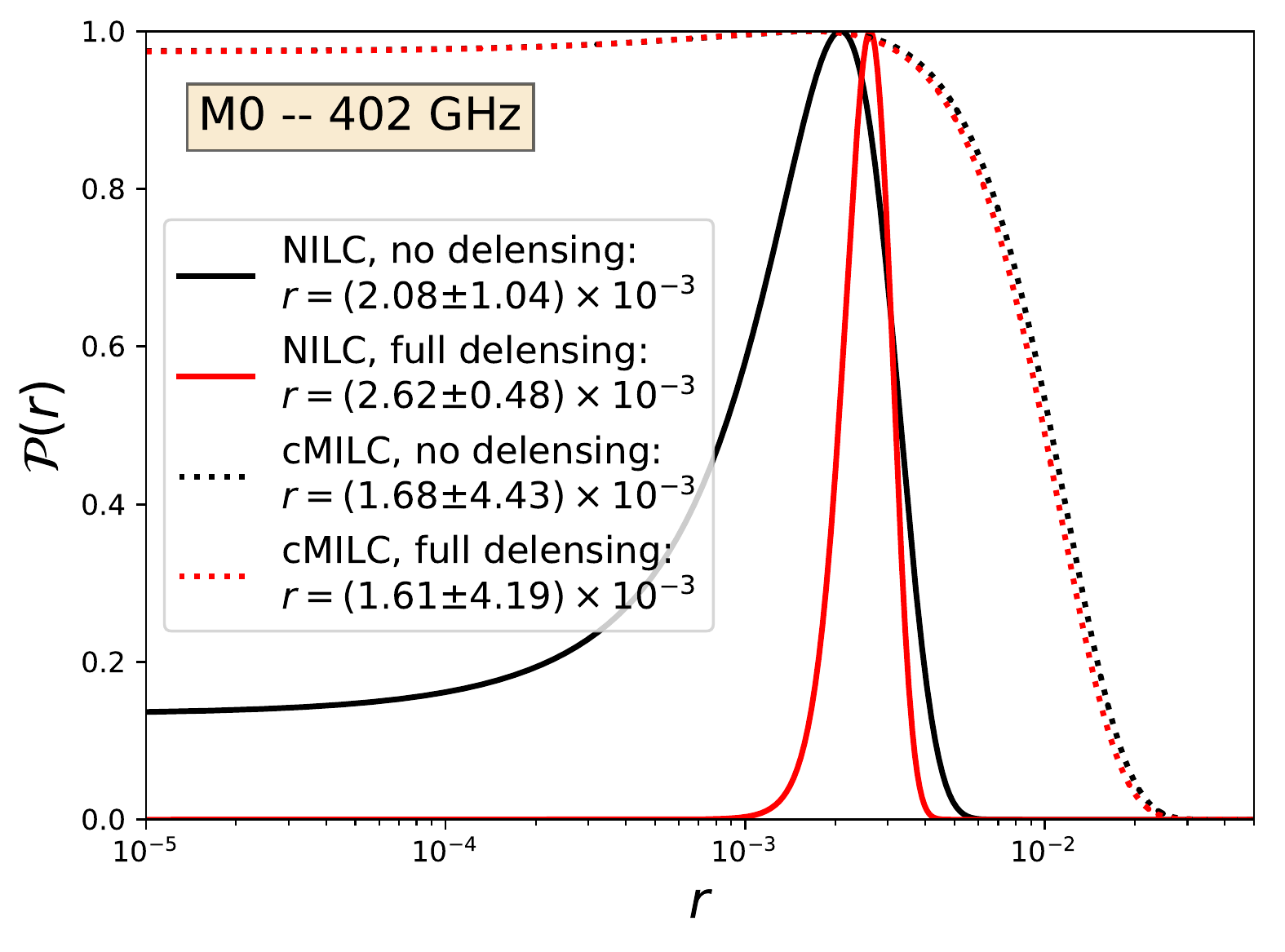}\\
    \includegraphics[width=\linewidth]{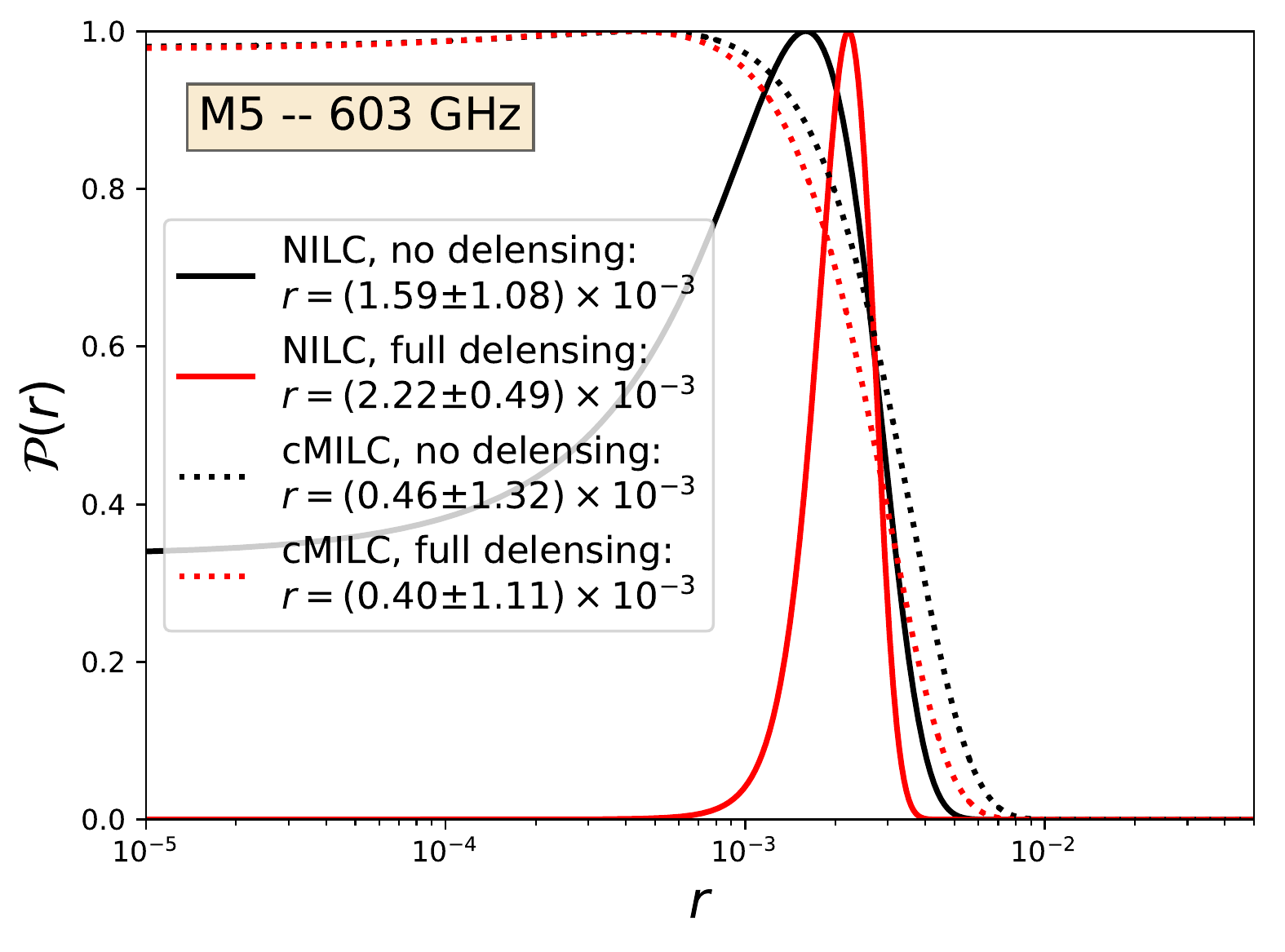}
  \end{center}
  \caption{Comparison of recovered tensor-to-scalar ratio
    distributions from \nilc\ (solid lines) and \cmilc\ (dotted lines)
    for the M0 (top panel) and M5 (bottom panel) instrument configurations. Results without
    delensing are shown as black curves, while results with full
    delensing are shown as red lines.}
  \label{fig:nilc_r}
\end{figure}

\begin{figure}
    \includegraphics[width=\linewidth]{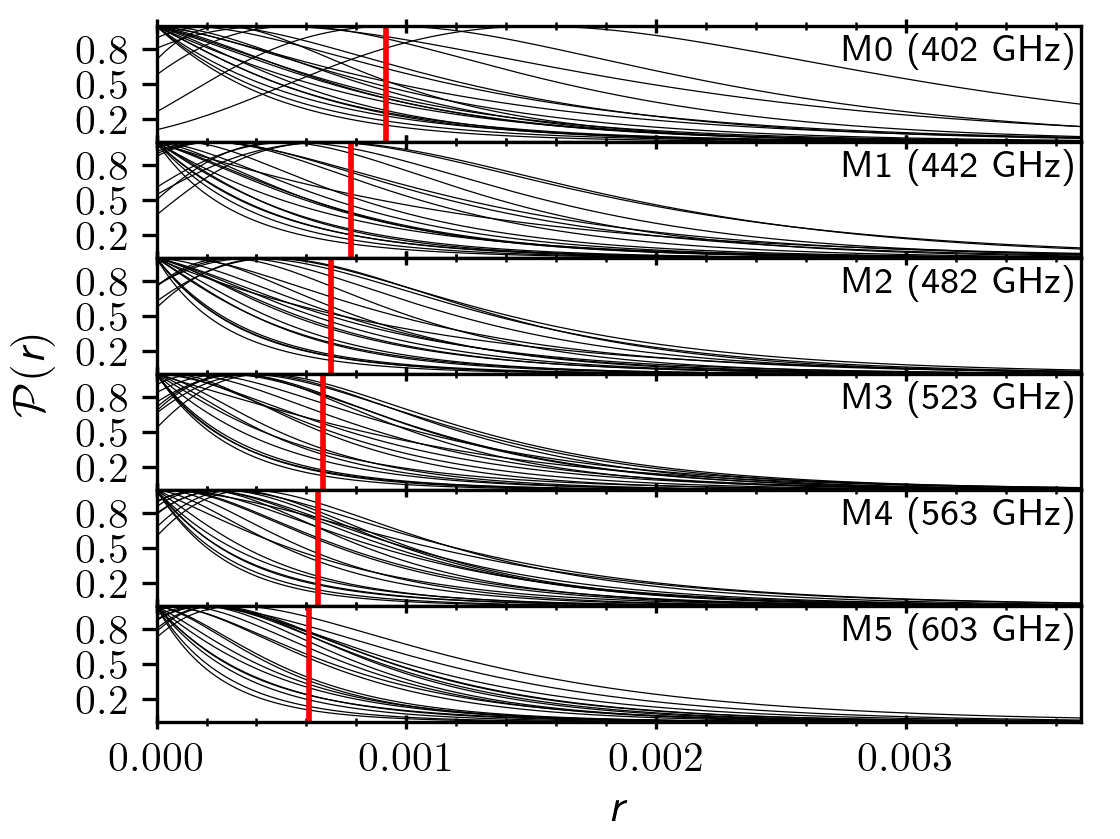}
  \caption{Tensor-to-scalar ratio posterior distributions for the single MBB component model (for both simulating and modeling) as derived with \commander\ using multipoles $\ell=[2,12]$ and a sky fraction of $f_{\mathrm{sky}}=73\,\%$ for 20 independent simulations (black lines), each drawn from a $\Lambda$CDM spectrum with $r=0$. The red vertical lines correspond to the mean upper $68\,\%$ confidence limit, $\sigma_r$.}
  \label{fig:comm_r_mbb}
\end{figure}

We now turn our attention to tensor-to-scalar ratio constraints, and
we start with probability distributions as derived
using the \nilc\ and \cmilc\ methods. These are summarized in
Fig.~\ref{fig:nilc_r}.\footnote{Although $r$ posterior plots are shown in terms of a logarithmic $x$-axis, the actual distributions are defined in linear units; this choice is made for plotting purposes only.} These distributions are derived by averaging over an ensemble of simulations, and therefore correspond inherently to ensemble averages. The top and bottom panels show the results for the
M0 and M5 configurations, respectively. Solid and dotted lines
show recovered tensor-to-scalar ratio distributions for \nilc\ and
\cmilc, while red and black lines show results with and without
delensing.

Since the fiducial value of the tensor-to-scalar ratio in these
simulations is $r=0$, any apparent bias ($r\neq 0$, defined by the
peak of the likelihood) is due to the power spectrum of the residual
foreground contamination that projects into the CMB $B$-mode maps
shown in Fig.~\ref{fig:nilc_map}. In contrast, the width of the
distribution, that is $\sigma_r$, receives contributions from the
cosmic variance of the lensed CMB signal, the sample variance of
projected foregrounds, and noise.

Starting with the unlensed \nilc\ results (solid black lines), we first note
that both M0 and M5 have a well-defined nonzero peak, which is
indicative of a nonnegligible foreground residual, consistent with
the nonnegligible foreground power seen in the left panel of
Fig.~\ref{fig:nilc_ps}. For the delensed case (red curve), this spurious
detection is statistically significant at the $5\,\sigma$ level
for M0, while for the lensed case it is nearly
consistent with zero at the $2\,\sigma$ level; however, this
difference is simply due to the additional uncertainty added by
lensing, and not lower foreground residuals as such. For M5, the
overall \nilc\ bias is reduced by about 30\,\% in both cases, while
the uncertainty remains unchanged, since this is dominated by cosmic
variance.

Being a more constrained version of the ILC, the semiblind
\cmilc\ method allows further foreground deprojection, and thereby
suppresses some of the residual biases that \nilc\ suffers
from. Specifically, by assuming an MBB ${f_{\rm
    dust}(\nu)=\nu^{\,\overline{\beta}_\mathrm{d}+1}/(e^{h\nu/k\overline{T}_\mathrm{d}}-1)}$
with pivots
${\overline{\beta}_\mathrm{d}=1.5,\overline{T}_\mathrm{d}=20\,{\rm
    K}}$ and a power-law ${f_{\rm
    sync}(\nu)=\nu^{\,\overline{\beta}_\mathrm{s}}}$
with pivot $\overline{\beta}_\mathrm{s}=-3$ as the baseline
zeroth-order SEDs for dust and synchrotron, \cmilc\ imposes four
nulling constraints to deproject the zeroth-order moments of dust and
synchrotron and the first-order moments of dust, whose respective SEDs
are $f_{\rm sync}(\nu)$, $f_{\rm dust}(\nu)$, $\partial f_{\rm
    dust}(\nu)/\partial\overline{\beta}_\mathrm{d}$, and $\partial f_{\rm
    dust}(\nu)/\partial\overline{T}_\mathrm{d}$.

The recovered distributions of the tensor-to-scalar ratio from
\cmilc\ are shown as dotted lines in Fig.~\ref{fig:nilc_r}. In
contrast to \nilc, \cmilc\ shows unbiased recovery of ${r=0}$ thanks
to deprojection of foreground moments. However, for M0 the extra
\cmilc\ constraints also significantly increase the noise
contribution to $\sigma_r$. On the other hand, for M5
\cmilc\ provides unbiased recovery of ${r=0}$, with a negligible
increase in noise. Thus, while M0 does not provide enough constraining
power to allow deprojection of thermal dust temperature moments at a
useful level, M5 does support this.

All the above results correspond to ensemble-averaged distributions
derived with blind internal linear combination methods. To understand
the behavior for individual realizations, Fig.~\ref{fig:comm_r_mbb}
shows corresponding posterior distributions as derived with \Commander\
for 20 realizations, plotted as thin lines for each of the six instrument
configurations. No delensing is applied in any of these cases, and a uniform prior on $r>0$ is assumed. The red
vertical lines show the mean of the upper $68\,\%$ confidence
limits. Firstly, we note that these distributions appear (at least
visually) statistically consistent with the input value of
$r=0$. Secondly, we also see that the upper limit decreases monotonically
with instrument configuration, with M5 providing the tightest overall
constraints. Thirdly, the internal scatter between individual
realizations is notably smaller with a higher maximum frequency,
resulting in an overall lower sample variance. 

\begin{figure}
  \begin{center}
    \includegraphics[width=\linewidth]{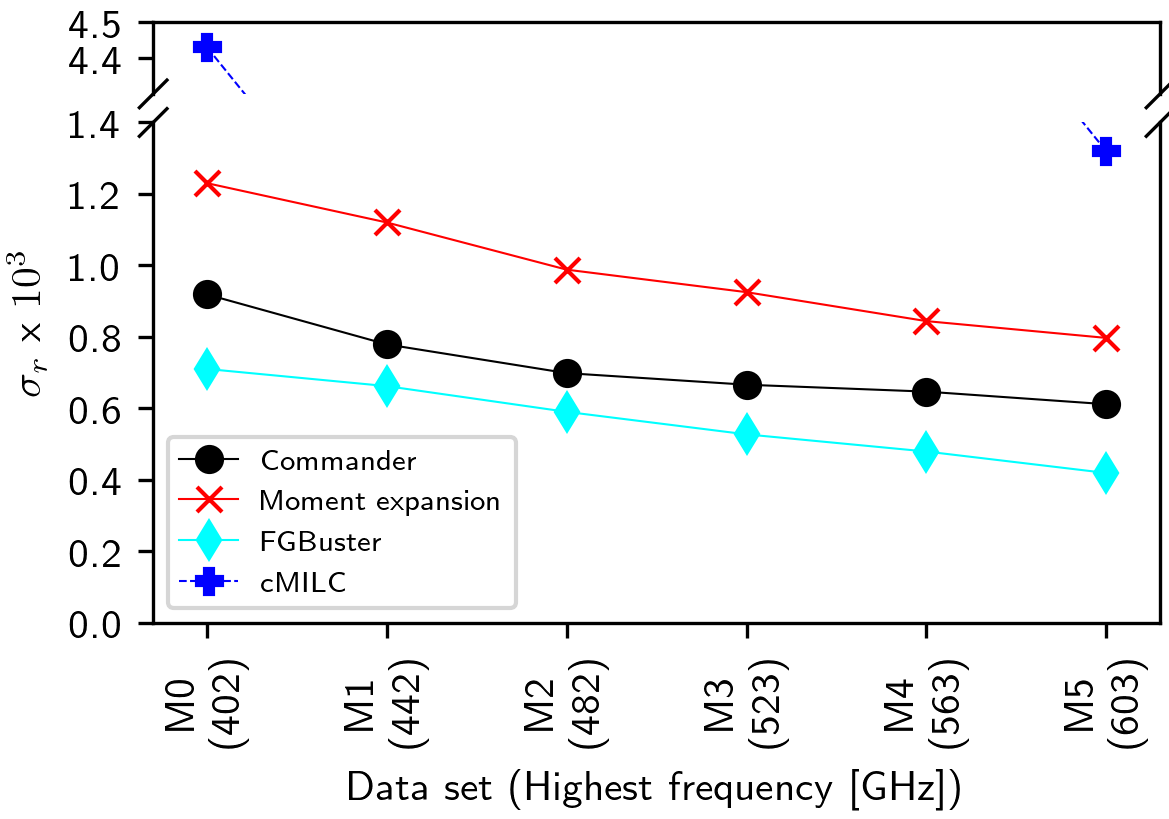}
  \end{center}
  \caption{Tensor-to-scalar ratio uncertainty, $\sigma_r$, as a
    function of instrument configuration for the
    component separation methods considered in this paper. }
  \label{fig:1mbb_allmethods}
\end{figure}

\begin{figure}
  \begin{center}
    \includegraphics[width=\linewidth]{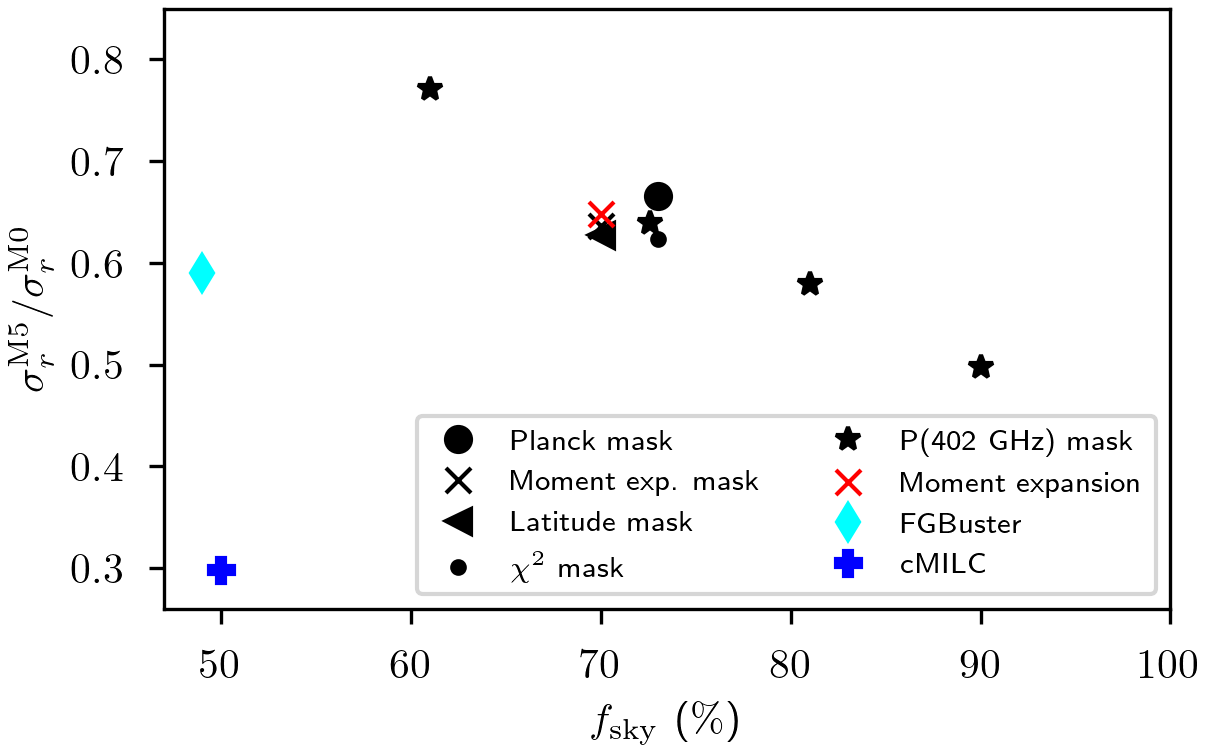}
  \end{center}
  \caption{Ratio between the tensor-to-scalar ratio uncertainties for
    M5 and M0 for different analysis choices and component-separation methods. The black points are \commander\ analyses using different masks and sky fractions.}
  \label{fig:masks}
\end{figure}

These general observations are summarized more quantitatively in
Fig.~\ref{fig:1mbb_allmethods}, which shows $\sigma_r$ as a function
of instrument configuration, as derived using the different 
component-separation methods. We note that $\sigma_r$ for \NILC\ is dominated by the cosmic variance of the bias, so this method is omitted from the figure. We note again that each of the methods involve
different masks, different multipole ranges, and different foreground
models, and the relative absolute shift between the curves is therefore
entirely expected. Furthermore, we also note that the baseline configuration does reach the target sensitivity of $\sigma_r < 0.001$ for both the \FGBuster\ and \commander\ methods, as already reported by \citet{PTEP}; even lower absolute uncertainties can be achieved by exploiting additional sky coverage (which is particularly relevant for \FGBuster) or multipole range (which  is particularly relevant for \commander).

More importantly for the main topic of this paper,
however, is the relative behavior between different configurations 
when we shift the entire HFT toward higher frequencies,
and this is qualitatively similar for all methods: the uncertainty on
$r$ decreases monotonically for all methods over the frequency range
considered in this paper. 

In Fig.~\ref{fig:masks} we plot the ratio
$\sigma_r^{\mathrm{M5}}/\sigma_r^{M0}$ for different sky fractions and
analysis methods. Overall, we see that M5 typically has 30--50\,\%
smaller uncertainties than M0, and the gains are larger when more sky
is included in the analysis. This makes intuitive sense, since accurate
component-separation is relatively more important near the Galactic
plane.

\begin{figure*}
\centering
\includegraphics[width=0.24\linewidth]{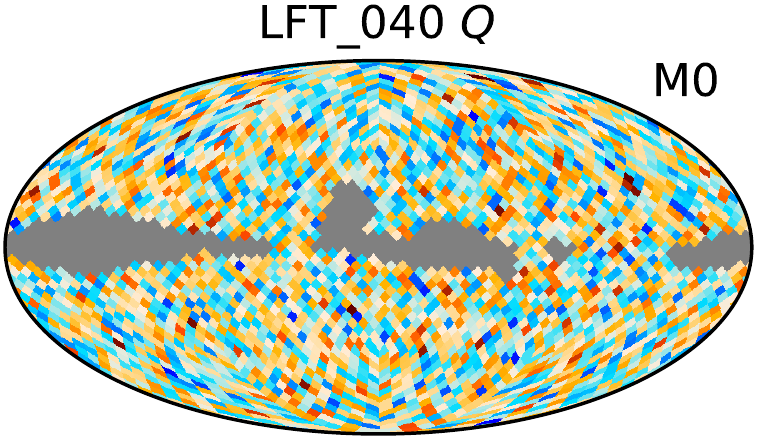}
\includegraphics[width=0.24\linewidth]{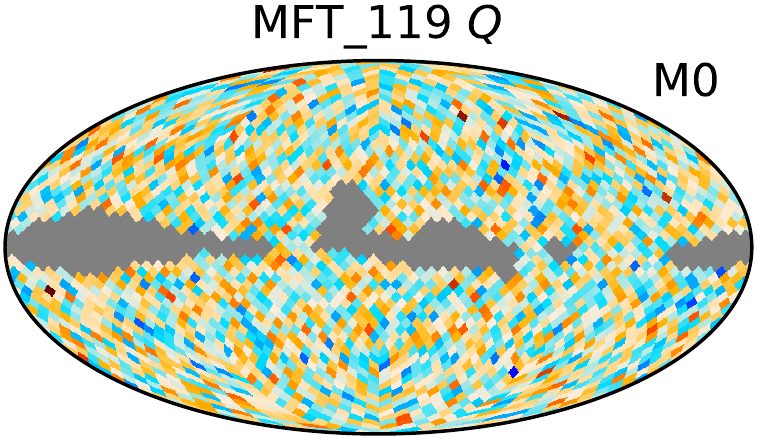}
\includegraphics[width=0.24\linewidth]{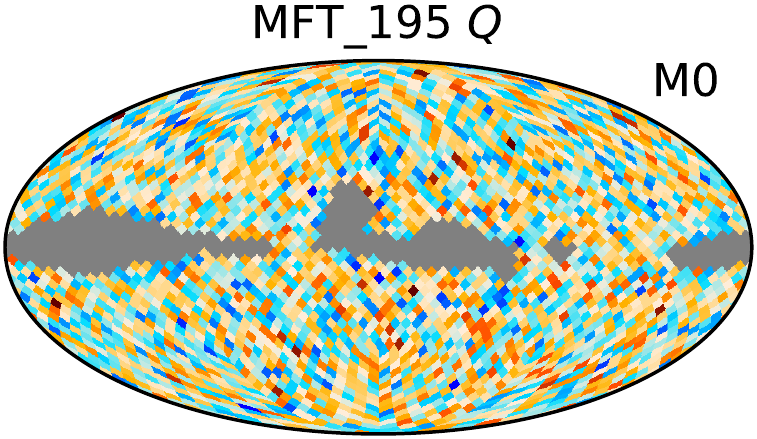}
\includegraphics[width=0.24\linewidth]{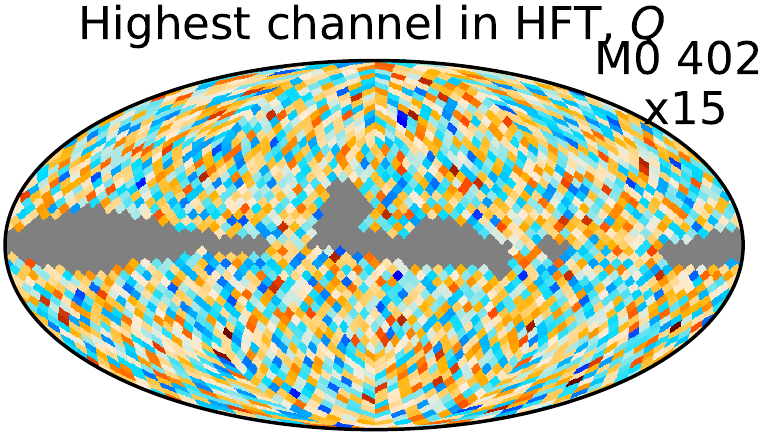}
\includegraphics[width=0.24\linewidth]{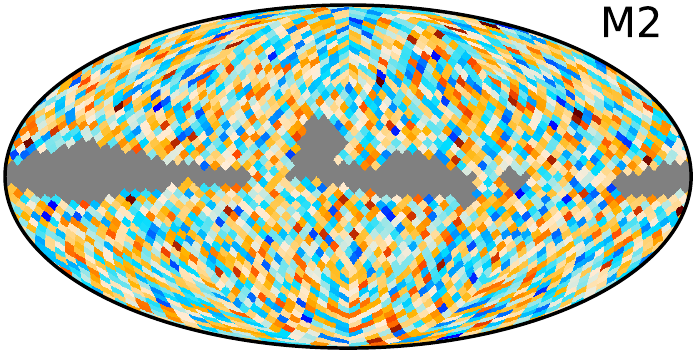}
\includegraphics[width=0.24\linewidth]{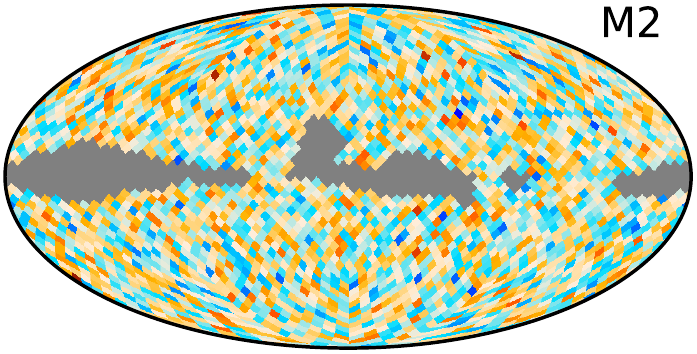}
\includegraphics[width=0.24\linewidth]{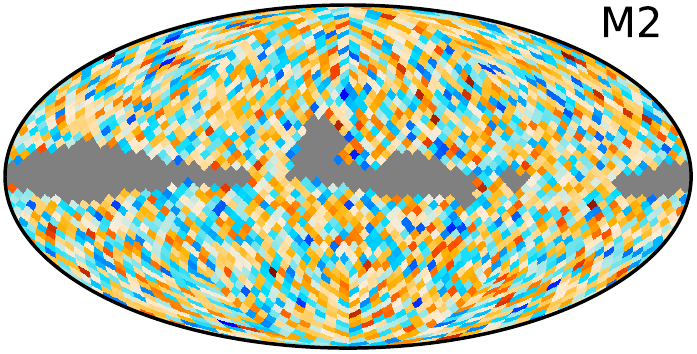}
\includegraphics[width=0.24\linewidth]{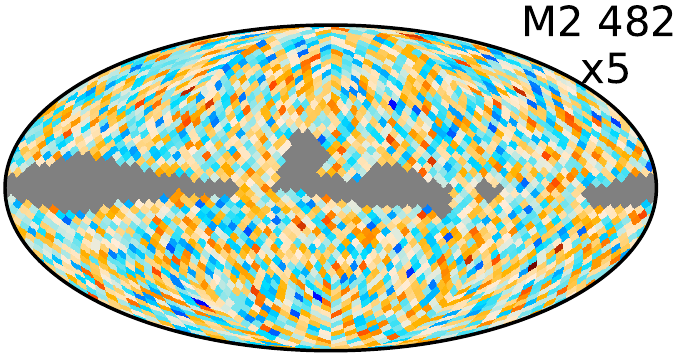}
\includegraphics[width=0.24\linewidth]{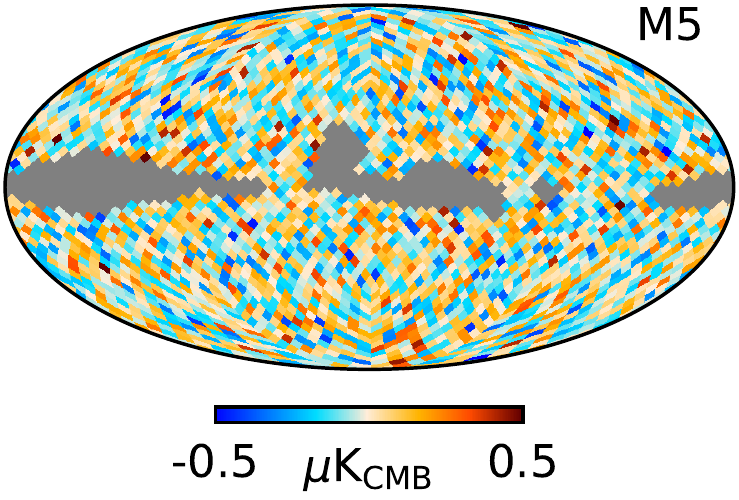}
\includegraphics[width=0.24\linewidth]{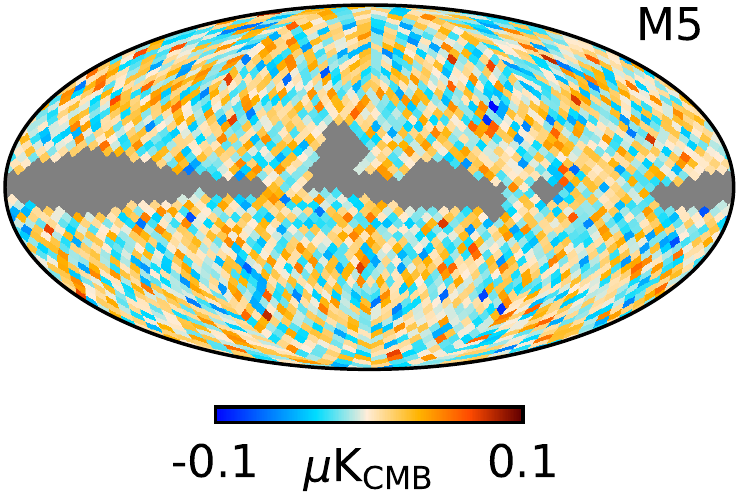}
\includegraphics[width=0.24\linewidth]{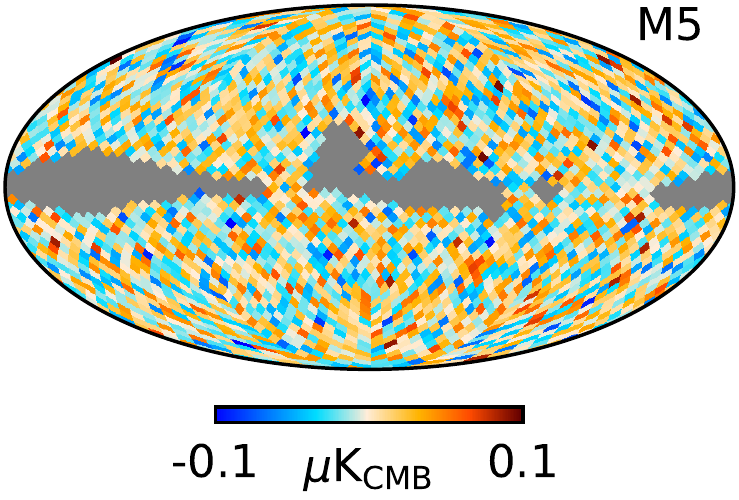}
\vspace{0.5cm}  
\includegraphics[width=0.24\linewidth]{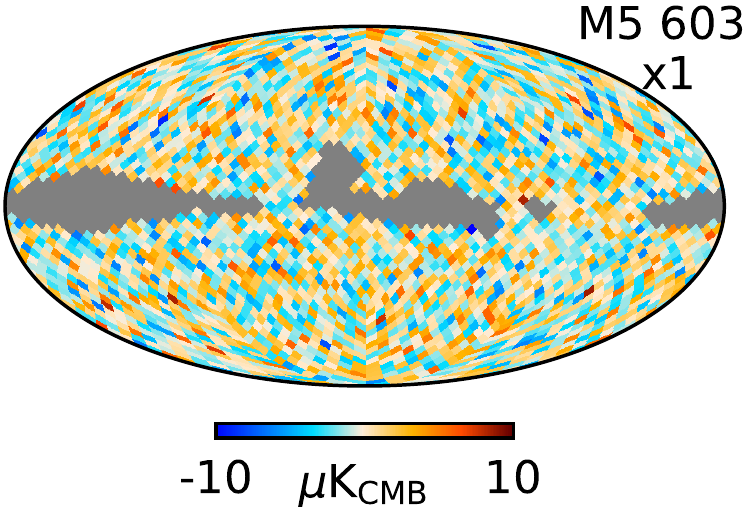}
\includegraphics[width=0.24\linewidth]{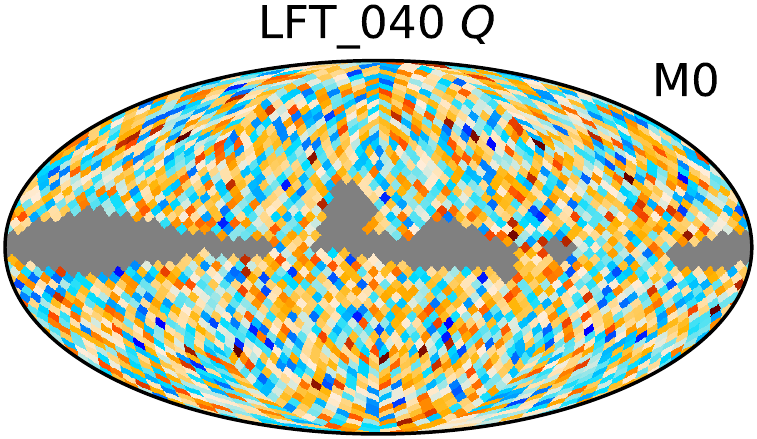}
\includegraphics[width=0.24\linewidth]{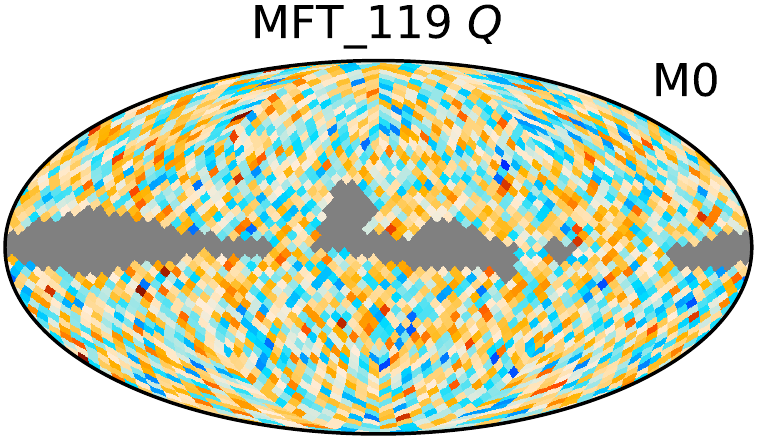}
\includegraphics[width=0.24\linewidth]{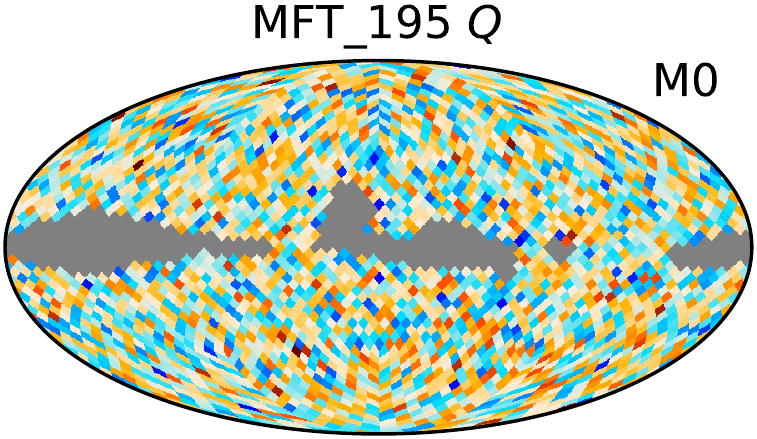}
\includegraphics[width=0.24\linewidth]{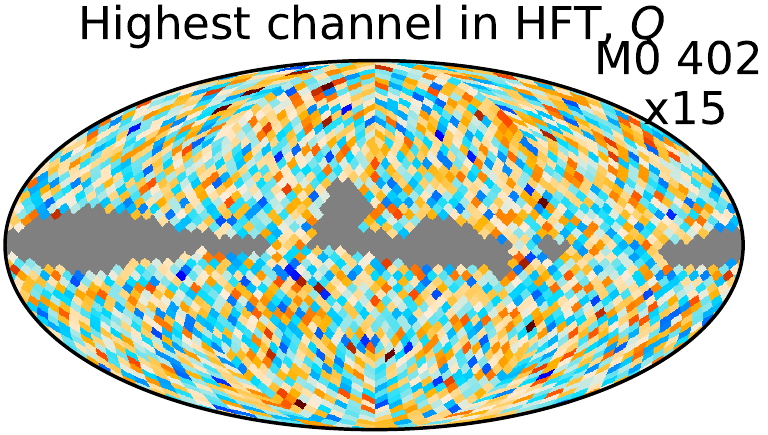}
\includegraphics[width=0.24\linewidth]{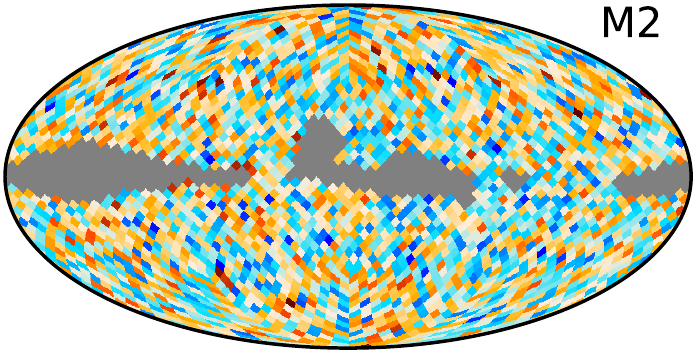}
\includegraphics[width=0.24\linewidth]{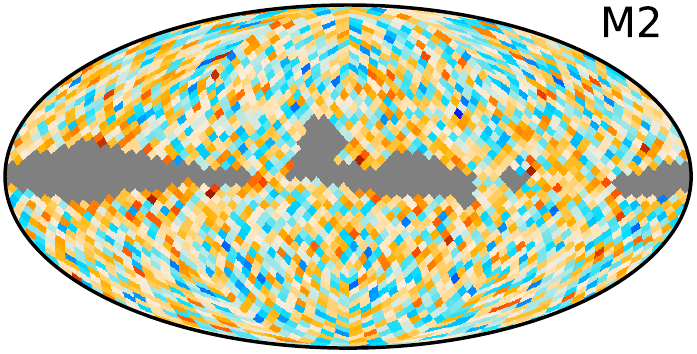}
\includegraphics[width=0.24\linewidth]{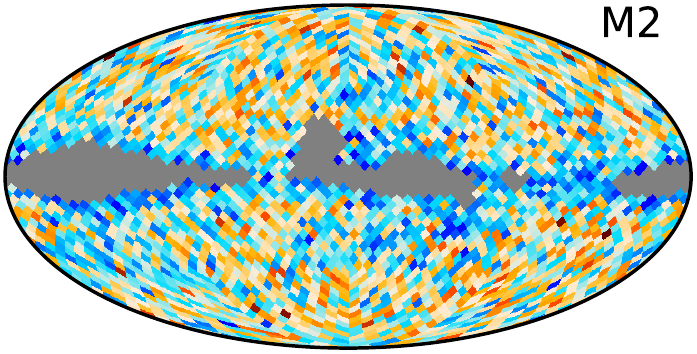}
\includegraphics[width=0.24\linewidth]{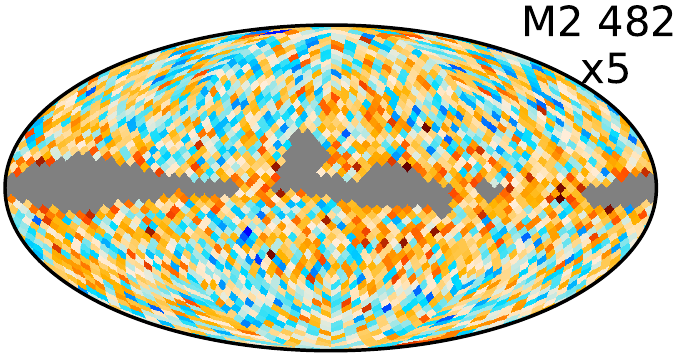}
\includegraphics[width=0.24\linewidth]{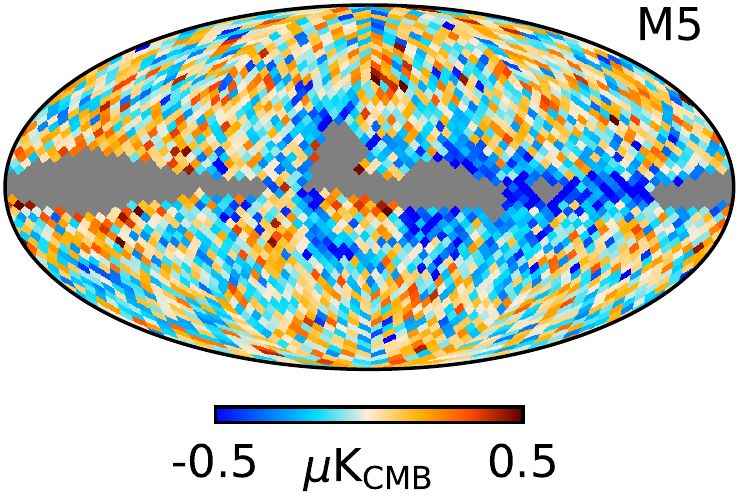}
\includegraphics[width=0.24\linewidth]{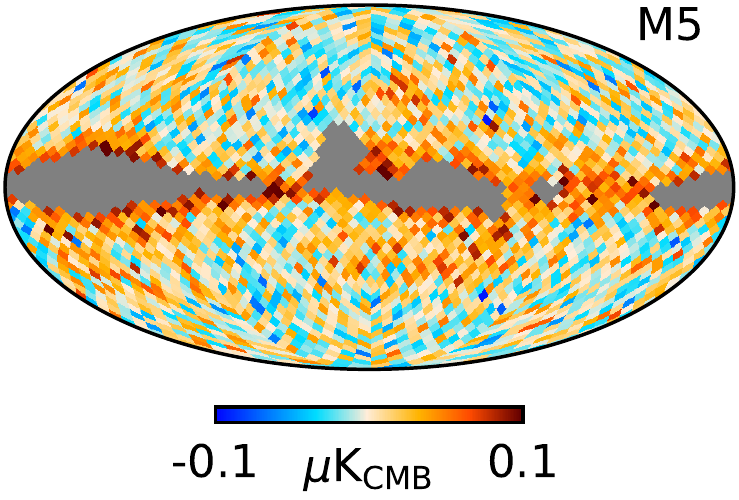}
\includegraphics[width=0.24\linewidth]{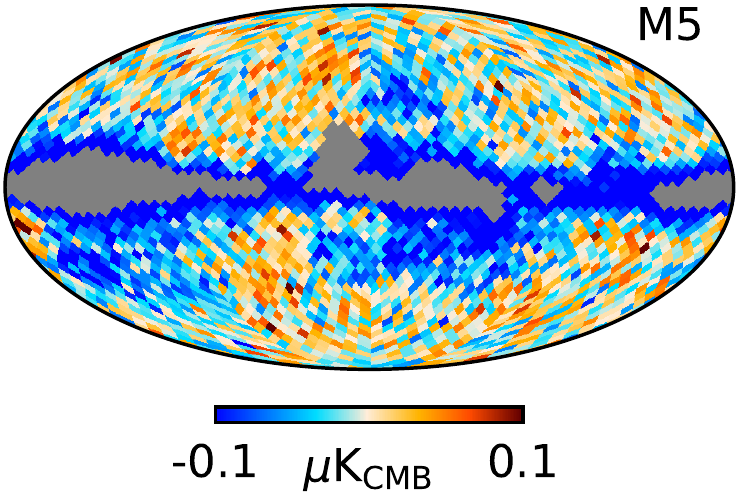}
\includegraphics[width=0.24\linewidth]{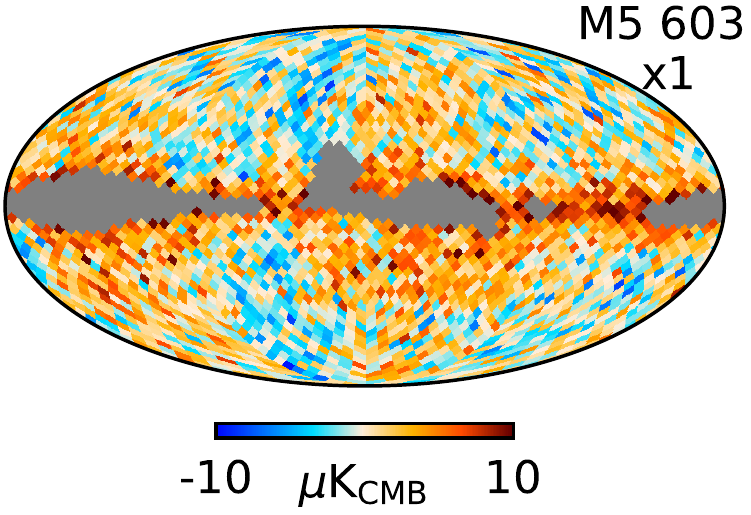}
\caption{\commander\ residual maps ($\r_{\nu}=\d_{\nu}-\s_{\nu}$) for four different frequency channels and three different instrument configurations. The top section show residual maps for the ideal one-component MBB model, while the bottom section show results for the nonideal SiFeC model, where a model mismatch was included.}
\label{fig:comm_residuals}
\end{figure*}

\subsection{Excluding the highest frequency channel}

As seen in the above calculations, extending the \litebird\ frequency
range from 402 to 603\,GHz leads to a substantially smaller
$\sigma_r$. However, as discussed in greater detail in the next
section, an extended frequency range is also associated with a
potentially larger bias from modeling errors, since the overall component
separation process becomes more dependent on assumptions regarding the
thermal dust SED, both in terms of its specific parametric shape and
its general spatial coherence between different frequency channels.

Recognizing these challenges, it is interesting to investigate how much the 
uncertainty on $r$ degrades if we exclude the highest frequency channel.
This represents the
scenario in which we are unable to model thermal dust at frequencies
above 500\,GHz. In this exercise, we adopt the M2
configuration, with the highest frequency channel at 482\,GHz, as a worked
example. When excluding the highest frequency channel in this particular
configuration, we recover an instrument model that is very similar in
frequency coverage to the baseline, M0, except with slightly lower
sensitivity in the remaining HFT channels.

Analyzing this truncated data set with \commander\ in the same manner
as those above, we find $\sigma_r = 9.8\times 10^{-4}$, which is
to be compared with $\sigma_r = 9.2\times 10^{-4}$ for the
nominal M0 configuration with otherwise identical settings, or an
increase of about 5\,\%. The potential loss in sensitivity is thus
relatively modest for this scenario. However, it is
important to note that an additional effect of this modification is
that there is no frequency overlap between the MFT and HFT in the
extended M2 configuration, while there is such overlap in the baseline
M0 configuration. Effectively, the extended instrument configurations
thus trade off control of unknown systematics in the overlap between
the MFT and HFT at the 195\,GHz band for additional sensitivity to
thermal dust emission at the highest frequencies.

\section{Non-ideal model analysis: Sensitivity to modeling errors}
\label{sec:complex_dust}

We now turn our attention to the issue of modeling errors, and we
consider two different types. The first type consists of SED modeling errors,
in which the fitted SED model is significantly different from the SED
of the true sky signal. Being able to reject a spurious
foreground-induced detection is key for a robust instrument
design. The second type arises from over-smoothing of spatially varying SED
parameters, which also can introduce biases.

\subsection{Detectability of SED model errors}

\begin{figure}
  \begin{center}
    \includegraphics[width=\linewidth]{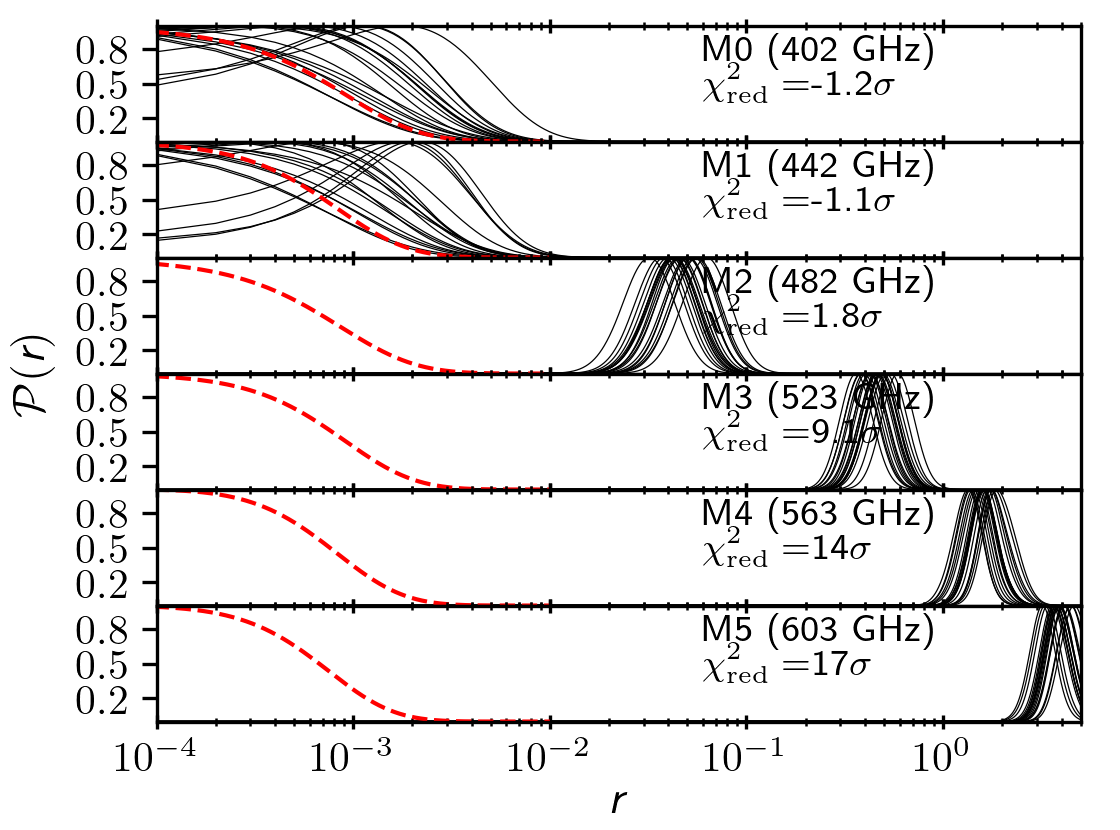} 
  \end{center}
  \caption{Tensor-to-scalar ratio posterior distributions with a fiducial tensor-to-scalar ratio of $r=0$, as derived with \commander\ when fitting SiFeC simulations using the one-component MBB model for six different instrument configurations. A $\chi^2$ statistic is reported for each 
  configuration, indicating the map-level goodness of fit. All results are derived from a sky fraction of $f_{\mathrm{sky}}=60\,\%$. The red dashed lines correspond to the median distribution from the ideal one-component MBB fit to an MBB simulation shown in Fig.~\ref{fig:comm_r_mbb} and serves as a reference.}
  \label{fig:comm_wrong_model}
\end{figure}

\begin{figure*}
  \begin{center}
    \includegraphics[width=\linewidth]{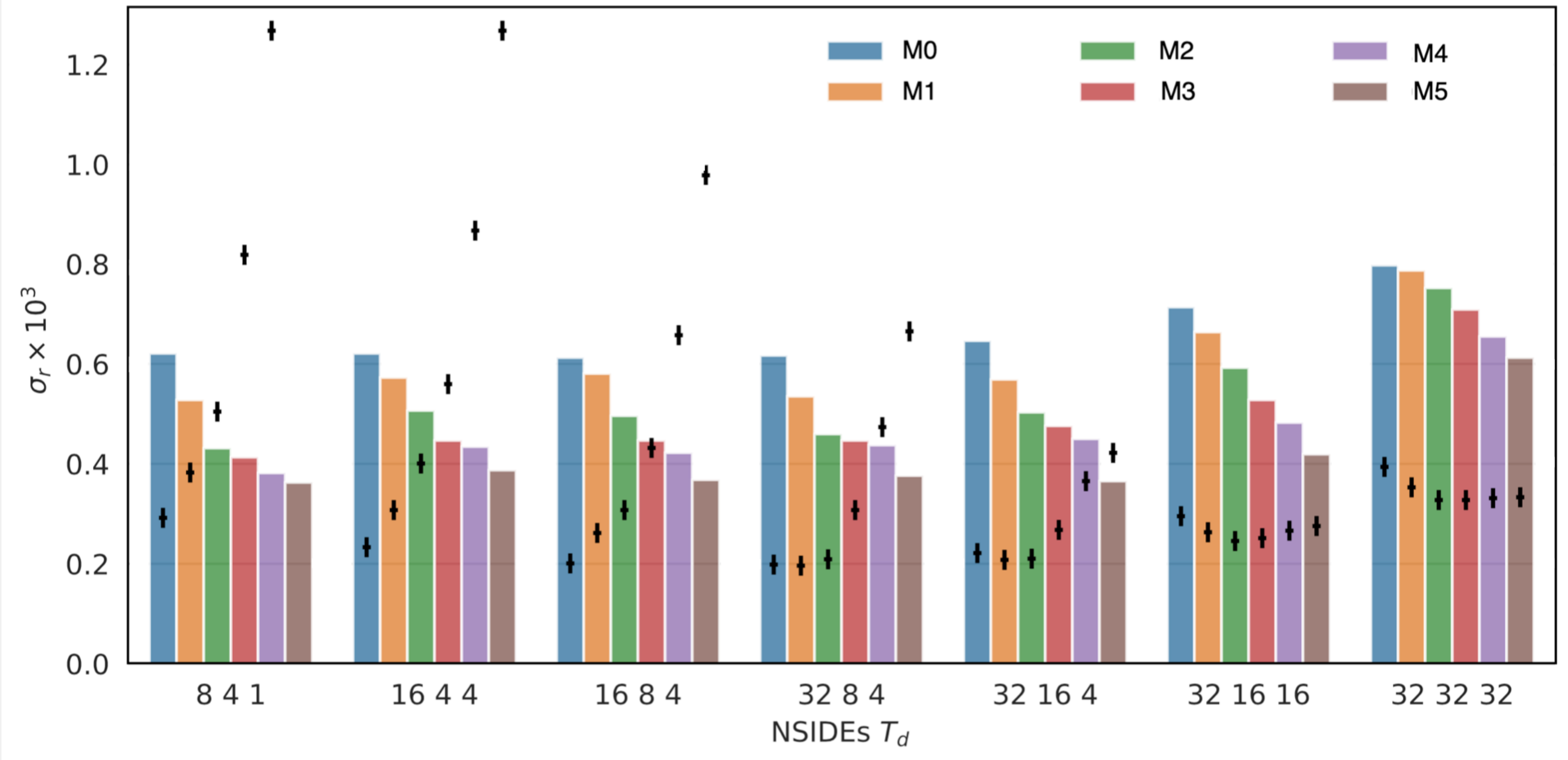}
  \end{center}
  \caption{
  Fisher uncertainty (bars) and bias (black crosses) on $r$ for different instrumental (colors) and \FGBuster\ component-separation configuration (bar groups). 
  The values on the horizontal axis define the $N_{\rm side}$ resolution at which the temperature of the thermal dust component is fitted for in the best 20\,\%, 20--40\,\% and 40--60\,\% sky fractions.
  }
  \label{fig:fgbuster_r_sigma}
\end{figure*}

To assess the ability to detect SED modeling errors as a function of
instrument configuration, we employ the parametric \Commander\ method,
which provides various goodness-of-fit quantities in terms of residual
maps and $\chi^2$ statistics among its output products. Specifically,
we analyze simulations based on the complex SiFeC
model summarized in Sect.~\ref{sec:simulations}, but using a simple
one-component MBB dust model for fitting. For each instrument
configuration and simulation, we compute both the tensor-to-scalar
ratio posterior distributions, as in Sect.~\ref{sec:r_constraints},
and the map-level reduced normalized $\chi^2$ statistic as
defined by
\begin{linenomath*} 
\begin{equation}
\chi^2_{\mathrm{red}} = \frac{\sum_{\nu,p} (d_{\nu,p}-s_{\nu,p})^2/\sigma_{\nu,p}^2 - n_{\mathrm{dof}}}{\sqrt{2n_{\mathrm{dof}}}}.
\end{equation}
\end{linenomath*} 
In this expression, the sum runs over all unmasked pixels, $p$, and
all frequencies, $\nu$, and $n_{\mathrm{dof}}$ is the total number of
degrees of freedom per pixel. Since the $\chi^2$ distribution
converges to a Gaussian with mean $n_{\mathrm{dof}}$ and variance
$2n_{\mathrm{dof}}$ for large $n_{\mathrm{dof}}$, this quantity
measures statistical outliers in units of standard deviations. In this
section, both the posterior distributions and the
$\chi^2_{\mathrm{\red}}$ statistics are evaluated on the unmasked
pixels encompassing $f_{\mathrm{sky}}=60\,\%$. However, we note that
the choice of sky fraction is arbitrary, as the main topic of this
section is the impact of modeling errors; a smaller sky fraction will
necessarily lead to smaller biases, but larger uncertainties, and vice
versa. The choice adopted here is simply chosen as a representative
compromise between minimizing both bias and uncertainties.

To build intuition regarding the following results, we show in the top
section of Fig.~\ref{fig:comm_residuals} residual maps ($\r_{\nu} =
\d_{\nu} - \s_{\nu}$) for one arbitrarily selected realization and
four frequency channels for the ideal one-component MBB model. In this
case we see no significant residuals in any frequency channels or
instrument configurations, simply because the fitted SED model matches
the actually simulated sky. All these maps are thus consistent
with white noise, and the corresponding $\chi^2$ statistic is also
consistent with the Gaussian noise expectation.

In the bottom section of Fig.~\ref{fig:comm_residuals}, we show
similar residual maps for the complex SiFeC simulations. Here we first
see that the baseline M0 configuration appears visually consistent
with the ideal MBB simulation in the top panel, and the parameter
estimation algorithm will clearly not be able to identify the
intrinsic model mismatch. Extending the frequency range to 482\,GHz (M2),
however, leaves noticeable residuals tracing high dust emission regions
in the Galactic plane, and at 603\,GHz (M5) the model mismatch is obvious.

The effect of these mismatches in terms of the tensor-to-scalar ratio
posterior distributions is quantified in
Fig.~\ref{fig:comm_wrong_model}. Each panel shows 20 realizations as
black lines, while the red line shows the median of the ideal
one-component MBB simulations as a reference; since the overall
foreground levels of the two sky models are comparable, and both the
fitted model and the instrumental noise parameters are identical
between the SiFeC and MBB models, the corresponding posterior
distributions should also be roughly equal.

Starting with the baseline configuration in the top panel, we see that
the tensor-to-scalar ratio is biased high by a factor of about two
(as seen in the shift between the black and red lines), even though
each distribution is individually consistent with zero. The $\chi^2$
statistic is completely consistent with the Gaussian hypothesis, with
a deviation of only $-1.2\,\sigma$. The M1 configuration behaves very
similarly.

For the case M2, we see that the bias in $r$ is greatly increased, to the point
that this configuration would have reported more than a $3\,\sigma$
detection of nonzero $B$-mode power. At this point, the $\chi^2$ value has
also increased somewhat, but is still only $2\,\sigma$ high compared
to the Gaussian expectation. It is only with frequency ranges equal
to or higher than 523\,GHz that $\chi^2$ is sufficiently deviant that 
we are able to conclusively detect the model mismatch, with a 
statistical significance of $9\,\sigma$ or more.

While the combination of a significant bias of $r$ and the statistical
acceptance of M2 appear disconcerting, it is important to recall that
these results are obtained with a relatively conservative sky fraction
of 60\,\%. For an analysis of real data, one would also exploit the
information in the high dust emission regions to identify a
statistically acceptable parametric model, and then fit that also at
high latitudes. As seen in the residual maps in
Fig.~\ref{fig:comm_residuals}, such an analysis would indicate that the
one-component MBB is sufficient for M0, while for M2 and higher
configurations the residuals are an obvious red flag. From a model
selection point of view, frequencies higher than $\approx\,$500\,GHz are
critically important for identifying poor thermal dust models.

\subsection{Resolution of spatially varying SED parameters}

Next, we consider the impact of different smoothing scales for SED
spectral parameters. In this case, we analyze the single MBB
simulations using the \FGBuster\ algorithm, which supports tuned pixel
sizes for each spectral parameter as part of its basic implementation,
quantified in terms of the {\tt HEALPix} resolution parameter,
$N_{\mathrm{side}}$. To disentangle the SED error effects from the
spatial resolution effect, we once again consider the ideal
one-component MBB simulations, but fit coarsely pixelized spectral
index maps, while the true spectral index maps are smoothly varying on
the sky. In particular, we focus on the thermal dust temperature,
which is key for the highest \LiteBIRD\ frequencies, and we allow this
to take on different resolutions at low, intermediate, and high
Galactic latitudes, denoted
$N_{\mathrm{side}}^{T_\mathrm{d}}=\{N_{\mathrm{side}}^\mathrm{low},
N_{\mathrm{side}}^\mathrm{int}, N_{\mathrm{side}}^\mathrm{high}\}$;
see Appendix~\ref{app:appendix_fgbuster} for details.

We are now interested in quantifying the recovery accuracy of $r$ as a
function of $N_{\mathrm{side}}^{T_\mathrm{d}}$, and the results from this analysis
are summarized in Fig.~\ref{fig:fgbuster_r_sigma} in terms of the
recovered mean tensor-to-scalar ratio $r$ (black crosses) and its
uncertainty (bars). Groups along the horizontal axis indicate
different $N_{\mathrm{side}}^{T_\mathrm{d}}$ combinations, while colors indicate
different \LiteBIRD\ configurations.

The first thing to note is that when the HFT channels are shifted to higher
frequencies, the uncertainty on the spectral indices decreases, and
this leads to a decrease of the statistical residuals, and therefore
also to an improvement in $\sigma_r$; this is the same effect as was
illustrated in Fig.~\ref{fig:1mbb_allmethods}, but now we also see
that this holds true almost independently of $N_{\mathrm{side}}^{T_\mathrm{d}}$. 

At the same time, we see that low values of 
$N_{\mathrm{side}}^{T_\mathrm{d}}$ lead to a strong bias in
$r$, and this effect increases rapidly with increasing frequency
range. On the other hand, this bias can be mitigated by increasing
$N_{\rm side}^{T_\mathrm{d}}$ at the expenses of an increase in the statistical
uncertainty. In particular, for $N_{\rm side}^{T_\mathrm{d}} =
\left[32, 32 ,32\right]$, all \LiteBIRD\ configurations lead to a bias
$\mathcal{O}(3\times 10^{-4})$, with an uncertainty $\sigma_r \approx
\mathcal{O}(6 \times 10^{-4})$.  In general, a maximum central
frequency above 500\,GHz requires at least 
$N_{\rm side}^{T_\mathrm{d}} = 16$, even for the cleanest 20\,\% of the sky 
in order to keep the bias under control.

\section{Discussion and conclusions}
\label{sec:conclusions}

One of the most important challenges for next-generation CMB
$B$-mode experiments is astrophysical foregrounds, and in particular
polarized thermal dust emission from our own Milky Way. Properly
choosing the frequency range is thus one of the key decisions to be
made for any given $B$-mode space mission or balloon-borne experiment. 
In this paper, we have revisited this issue for \LiteBIRD. Although the established baseline configuration has already been shown to achieve the main mission goals \citep{PTEP}, it is still desirable to optimize the overall uncertainties to increase the overall system margins.

In this paper, we have addressed this issue by analyzing simulations to measure the recovered uncertainty
on the tensor-to-scalar ratio using several different component-separation
algorithms for six different possible instrument
configurations, where the entire HFT frequency range is shifted
toward higher frequencies. We find that the total statistical 
uncertainty on the tensor-to-scalar ratio, $r$, after foreground cleaning
may be reduced by 30--50\,\% by extending
the current maximum frequency from the current baseline of 402\,GHz to
500\,GHz, or higher for the single MBB dust model, also
depending on the sky fraction. 

A wider frequency range also increases the
instrument's sensitivity to the shape of the foreground SED, and this
has two important consequences. On the one side, a wider frequency
range imposes stronger requirements on the foreground model itself, since
modeling errors may more easily contaminate the CMB results. For
instance, if a given component-separation algorithm (explicitly or
implicitly) assumes that the high-frequency channels correlate
perfectly with the thermal dust emission at lower frequencies, while
the true sky exhibits frequency decorrelation due to the 3D structure
of the Milky Way, some fraction of the difference between the assumed and
real sky model will leak into the CMB \citep{tassis2015}. It is therefore essential that
the method of choice allows for sufficient flexibility to capture such
uncertainties and model mismatches.

At the same time, a wider frequency range also provides the tools for
actually identifying and constraining that model in the first
place. Indeed, one of the most important concerns regarding a limited
frequency range is the possibility of having low-level foreground
residuals that mildly bias cosmological parameters, but still
result in an acceptable goodness-of-fit, (for instance as measured by
a $\chi^2$ statistic). Higher frequency channels provide important
safe-guards against this type of error. Additionally, having access
to higher frequency information increases the number of basis
functions that can be used to model the thermal dust emission as a
function of frequencies. A particularly important example of such a
mode is the spatial distribution of the thermal dust temperature,
which is useful for both blind and nonblind methods.

An important intuition that underlies all the results discussed in
this paper is simply the fact that the CMB SED falls nearly
exponentially (relative to dust) above 300\,GHz. This implies that even
relatively small modifications of the frequency range can lead to
significant improvements in a given instrument's ability to separate
CMB from thermal dust. In particular, at the sensitivity level of
\LiteBIRD, a 402\,GHz channel has a non-negligible signal-to-noise
ratio for CMB fluctuations, while at 500\,GHz this is essentially
zero. For an extended configuration, the highest frequency channel therefore
represents essentially a pure thermal dust map, while for the baseline
configuration it represents a weighted sum of CMB and thermal dust
emission. As shown quantitatively in this paper, a clean foreground
tracer substantially improves tensor-to-scalar constraints for both
parametric and nonparametric (ILC) component-separation methods. 

In this paper we have primarily addressed the impact of Galactic foregrounds from the point of view of CMB confusion. An important lesson learned from both \WMAP\ and \Planck, however, is that Galactic foregrounds also play a key role in understanding and mitigating low-level instrumental systematic effects. For instance,  confusion between transmission imbalance factors and foregrounds turned out to be an important uncertainty for \WMAP\ \citep{page2007,watts:2022}, while confusion between gain calibration and foregrounds was a key limitation for \Planck\ LFI \citep{PlanckI:2018all,beyondplanck2020,gjerlow2022}. One may therefore argue that experience shows that the most robust approach to mitigating the full effect of astrophysical foregrounds is not by avoiding them, but rather by measuring them. In addition to improving CMB constraints, a wide frequency range will also increase the amount of Galactic science that may be derived from these observations, and it will thereby also increase the overall legacy value of the mission for other fields of astronomy. 

While the various instrumental setups studied in this paper appear as possible options, we would like to stress that their technical feasibility has not been performed in depth yet. This would have to include both aspects of optical design and detection chains, but also the potential impacts on the needs for optical modeling and calibration facilities. This trade-off study would also have to deal with the impact of such modifications of the HFT design on the overall systematics effects, which should be propagated into systematics uncertainties on $r$. As noted in the introduction, performing such a  full trade-off study lies far beyond the scope of this paper. Rather, the goal of this paper is to understand whether sufficiently significant gains are achievable under ideal conditions to motivate a proper detailed study -- and with typical gains at the 30--50\,\% level in the statistical uncertainty on $r$ after foreground cleaning, the results do appear interesting enough to be investigated further.

\begin{acknowledgements}
%
This work is supported in Japan by ISAS/JAXA for Pre-Phase A2 studies, by the acceleration program of JAXA research and development directorate, by the World Premier International Research Center Initiative (WPI) of MEXT, by the JSPS Core-to-Core Program of A. Advanced Research Networks, and by JSPS KAKENHI Grant Numbers JP15H05891, JP17H01115, and JP17H01125. 
The Canadian contribution is supported by the Canadian Space Agency.
The French \LiteBIRD\ phase A contribution is supported by the Centre National d’Etudes Spatiale (CNES), by the Centre National de la Recherche Scientifique (CNRS), and by the Commissariat à l’Energie Atomique (CEA).
The German participation in \LiteBIRD\ is supported in part by the Excellence Cluster ORIGINS, which is funded by the Deutsche Forschungsgemeinschaft (DFG, German Research Foundation) under Germany’s Excellence Strategy (Grant No. EXC-2094 - 390783311).
The Italian \LiteBIRD\ phase A contribution is supported by the Italian Space Agency (ASI Grants No. 2020-9-HH.0 and 2016-24-H.1-2018), the National Institute for Nuclear Physics (INFN) and the National Institute for Astrophysics (INAF).
Norwegian participation in \LiteBIRD\ is supported by the Research Council of Norway (Grant No. 263011) and has received funding from the European Research Council (ERC) under the Horizon 2020 Research and Innovation Programme (Grant agreement No. 772253 and 819478).
The Spanish \LiteBIRD\ phase A contribution is supported by the Spanish Agencia Estatal de Investigación (AEI), project refs. PID2019-110610RB-C21,  PID2020-120514GB-I00, ProID2020010108 and ICTP20210008.
Funds that support contributions from Sweden come from the Swedish National Space Agency (SNSA/Rymdstyrelsen) and the Swedish Research Council (Reg. no. 2019-03959).
The US contribution is supported by NASA grant no. 80NSSC18K0132.
%
We also acknowledge funding from the European Research Council (ERC) under the Horizon 2020 Research and Innovation Programme (Grant agreement No.~725456 and 849169) and The Royal Society (Grant No.~URF/R/191023)

\end{acknowledgements}

\bibliographystyle{aa}

\bibliography{bibliography}

\begin{appendix}
\section{Parametric component-separation methods}
\label{app:parametric}

\subsection{Commander}
\label{app:appendix_comm}

\label{sec:comm_method}
\commander\ is a standard parametric Bayesian component-separation framework for CMB observations \citep{eriksen:2004,eriksen:2008} that has been used extensively by \Planck\ \citep[e.g.,][]{Planck2013compsep,Planck2016compsep,Planck2018compsep,npipe}. In the current analysis we employ the first implementation of this framework, typically referred to as \commanderone, which was used in the \Planck\ 2013 data release. The main advantage of this framework is a very high computational efficiency, which is important when analyzing an ensemble of simulations; at the same time, the main drawback is its requirement of identical angular resolution of all frequency bands, which limits its use to low-resolution data sets. In our current analysis, all input maps are at $N_{\mathrm{side}}=16$, with a common resolution of $10^\circ$ FWHM.
Another option would be to use \commanderthree\ \citep{beyondplanck2020}, which supports both multiresolution component separation and low-level time-ordered data analysis, but this would increase the computational expense of the analysis by orders of magnitude, without actually providing valuable new information regarding the central question of frequency range considered in this paper.

The starting point of any parametric Bayesian analysis is to define an explicit parametric data model, described in Eqs.~(\ref{eq:data_model})--~(\ref{eq:data_model_expanded}), and in \commanderone, every parameter is fitted per pixel. 
The main goal is now to fit the parameters $\theta = \{\vec{a}_{\mathrm{CMB}}, \vec{a}_{\mathrm{s}}, \vec{a}_\mathrm{d}, \beta_{\mathrm{s}}, \beta_\mathrm{d}, T_\mathrm{d}$\}, which within a Bayesian framework means computing the posterior distribution $P(\theta | \d )$, and then using this to estimate the cosmological parameter $r$. Using Bayes' theorem, the posterior distribution may be written as
\begin{linenomath*} 
\begin{align}
    P(\theta | \d ) & = \frac{P(\d | \theta )P(\theta)}{P(\d)} \\
    &\propto P(\d | \theta )P(\theta), \label{eq:bayes_theorem}
\end{align}
\end{linenomath*} 
where $P(\d)$ is a normalization factor, $P(\d | \theta )$ is the likelihood, and $P(\theta)$ represents some set of priors. In this work, we apply loose Gaussian priors on the spectral parameters, but no priors on the amplitude parameters. Specifically, we adopt a synchrotron prior of $P(\beta_{\mathrm{s}}) = N(-3.0, 0.3)$, where $N(\mu,\sigma)$ denotes a standard Gaussian distribution with mean $\mu$ and standard deviation $\sigma$. For thermal dust emission, we adopt $P(\beta_{\mathrm{d}}) = N(1.54, 0.20)$ and $P(T_{\mathrm{d}}) = N(23\,\mathrm{K}, 7\,\mathrm{K})$ where we fit for the one MBB component.

To map out the distribution in Eq.~(\ref{eq:bayes_theorem}), we use Gibbs sampling as implemented in \commander. That is, rather than sampling directly from $P(\theta | \d )$, which is computationally unfeasible, we iteratively sample from each of the corresponding conditional probability distributions. For the purposes of this paper, this can be written as
\begin{linenomath*} 
\begin{align}
    \{\vec{a}_{\mathrm{CMB}}, \vec{a}_{\mathrm{s}}, \vec{a}_\mathrm{d}\}^{i+1} &\leftarrow P(\vec{a}_{\mathrm{CMB}}, \vec{a}_{\mathrm{s}}, \vec{a}_{\mathrm{d}} | \beta_\mathrm{s}^i, \beta_\mathrm{d}^i, T_{\mathrm{d}}^i, \d), \\\label{eq:comm_amp}
    \{\beta_{\mathrm{s}}, \beta_{\mathrm{d}}, T_{\mathrm{d}}\}^{i+1} &\leftarrow P( \beta_{\mathrm{s}}, \beta_{\mathrm{d}}, T_{\mathrm{d}} | \vec{a}_{\mathrm{CMB}}^{i+1}, \vec{a}_{\mathrm{s}}^{i+1}, \vec{a}_\mathrm{d}^{i+1}, \d).
\end{align}
\end{linenomath*} 
Here $\leftarrow$ indicates that the parameters to the left are sampled from the distribution on the right-hand side. All amplitude parameters are sampled jointly together with the CMB sky signal to avoid excessive Monte Carlo correlation lengths. The nonlinear parameters, the spectral indices and dust temperature, are sampled using a standard inversion sampler.

We run this iterative Gibbs sampler until convergence, which in the current analysis requires about $40\,000$ samples after rejecting a burn-in period of about 500 samples. We repeat this analysis for 20 different CMB and noise realizations.\footnote{This large number of samples is the main motivation for using \commanderone\ rather than \commanderthree; producing $\mathcal{O}(10^6)$ posterior samples at full angular resolution would require tens of millions of CPU hours. While feasible for a final production analysis of real data, such a cost is not justified for the current exploratory analysis.} Based on the samples from each simulation, we compute a posterior mean CMB map, $\hat{\vec{a}}_{\mathrm{CMB}}$, and the corresponding covariance matrix, $\tens{N}$, and these are subsequently fed into a standard Gaussian likelihood;
\begin{linenomath*} 
\begin{equation}
\mathcal{L}(r) \propto \frac{e^{-\frac{1}{2} \hat{\vec{a}}_{\mathrm{CMB}}^t 
\left(\tens{S}(r) +
\tens{N}\right)^{-1}\hat{\vec{a}}_{\mathrm{CMB}}}}{\sqrt{|\tens{S}(r) +
\tens{N}|}}.
\label{eq:rlnL}
\end{equation}
\end{linenomath*} 
Here $\tens{S}(r)$ is the CMB signal covariance matrix for a given
value of $r$ and the best-fit $\Lambda$CDM parameters described in
Sect.~\ref{sec:simulations}. The low pixel resolution of
$N_{\mathrm{side}}=16$ used for \commander\ in this paper is dictated
by the CPU time and RAM required to evaluate this equation, as well as
the number of samples needed to establish a robust noise covariance
matrix. 

We adopt a uniform prior for all positive values of $r$, and the posterior distribution is therefore numerically identical to the likelihood in Eq.~(\ref{eq:rlnL}) for $r>0$. This function is mapped out by gridding Eq.~(\ref{eq:rlnL}) over a precomputed library of angular power spectra, $C_{\ell}(r)$, which are computed into covariance matrices in pixel space, $\tens{S}(r)$, including only multipoles in the range $\ell=[2,12]$, as dictated by the low pixel resolution of $N_{\mathrm{side}}=16$. The \commander\ results thus only probe the low-$\ell$ reionization peak of the $B$-mode power spectrum, not the recombination peak around $\ell\approx100$. However, this is also where diffuse component separation is most challenging, and the question of frequency optimization is most important.

The main analysis mask used for the \commander\ analysis is the $f_{\mathrm{sky}}=73\,\%$ polarization mask discussed in \citet{Planck2018compsep}. 
To visualize the impact of sky fraction, other masks have been used in Fig.~\ref{fig:masks}, such as masks based on high latitude cuts, $\chi^2$, and by thresholding the 402\,GHz polarization amplitude ($P = \sqrt{Q^2 + U^2}$) map, which primarily suppresses thermal dust emission.
This latter is used for the 60\,\% mask in  Sect.~\ref{sec:complex_dust}.

\subsection{FGBuster}
\label{app:appendix_fgbuster}
\label{sec:fgbuster_method}

A second implementation of parametric component separation is called 
\FGBuster.\footnote{\url{https://github.com/fgbuster/fgbuster}}
This method is based on the same parametric modeling as defined by Eqs.~\eqref{eq:data_model}--\eqref{eq:data_model_expanded}, but uses a computationally efficient non-linear optimization procedure to perform the fit, as opposed to brute-force Monte Carlo sampling as implemented in \commander.
Following~\cite{Stompor2009}, \cite{Errard2011}, \cite{Errard2018}, and \cite{Puglisi2022},
\FGBuster\ proceeds in three steps. These are: first the estimation of the spectral parameters $\{\beta_\mathrm{d}, T_\mathrm{d}, \beta_\mathrm{s}\}$ through the optimization of a so-called spectral likelihood; second, the estimation of the component maps $\{\vec{a}_{\mathrm{CMB}},\vec{a}_{\mathrm{d}},\vec{a}_{\mathrm{s}}\}$ introduced in Appendix~\ref{sec:comm_method}; and third, the estimation of the angular power spectra for the residual foregrounds (i.e., the difference between the input and recovered CMB map), as well as the estimation of the tensor-to-scalar ratio $r$ with a Gaussian likelihood defined by the previous quantities, the theoretical $BB$ spectrum, and the noise after component separation. 

Regarding the first step, we note that \FGBuster\ supports different resolutions (as defined by the {\tt HEALPix} resolution parameter, $N_{\mathrm{side}}$) for each spectral parameter, while the corresponding amplitudes are fitted at full angular resolution. Choosing a low $N_{\rm side}$ implies that few foreground parameters are fitted for, resulting in low post-component-separation noise; however, this may also result in a high bias due to sky complexity that the model is not allowed to capture. Optimizing $N_{\mathrm{side}}$ for each parameter is an important point of the algorithm, and we quantify these two competing effects by splitting the foreground residuals into two categories. First, the systematic residuals, as obtained from a noiseless simulation, are driven by the mismatch between input sky and the assumed parametric model, and its power spectrum, $B_\ell$, defines the bias on $r$. Second, the statistical residuals are computed as the average power spectrum from 10 noisy simulations, $S_{\ell}$, which are driven by the statistical error while estimating spectral parameters, and by coaddition of the noise at different frequencies.

\begin{figure}
  \begin{center}
    \includegraphics[width=\linewidth]{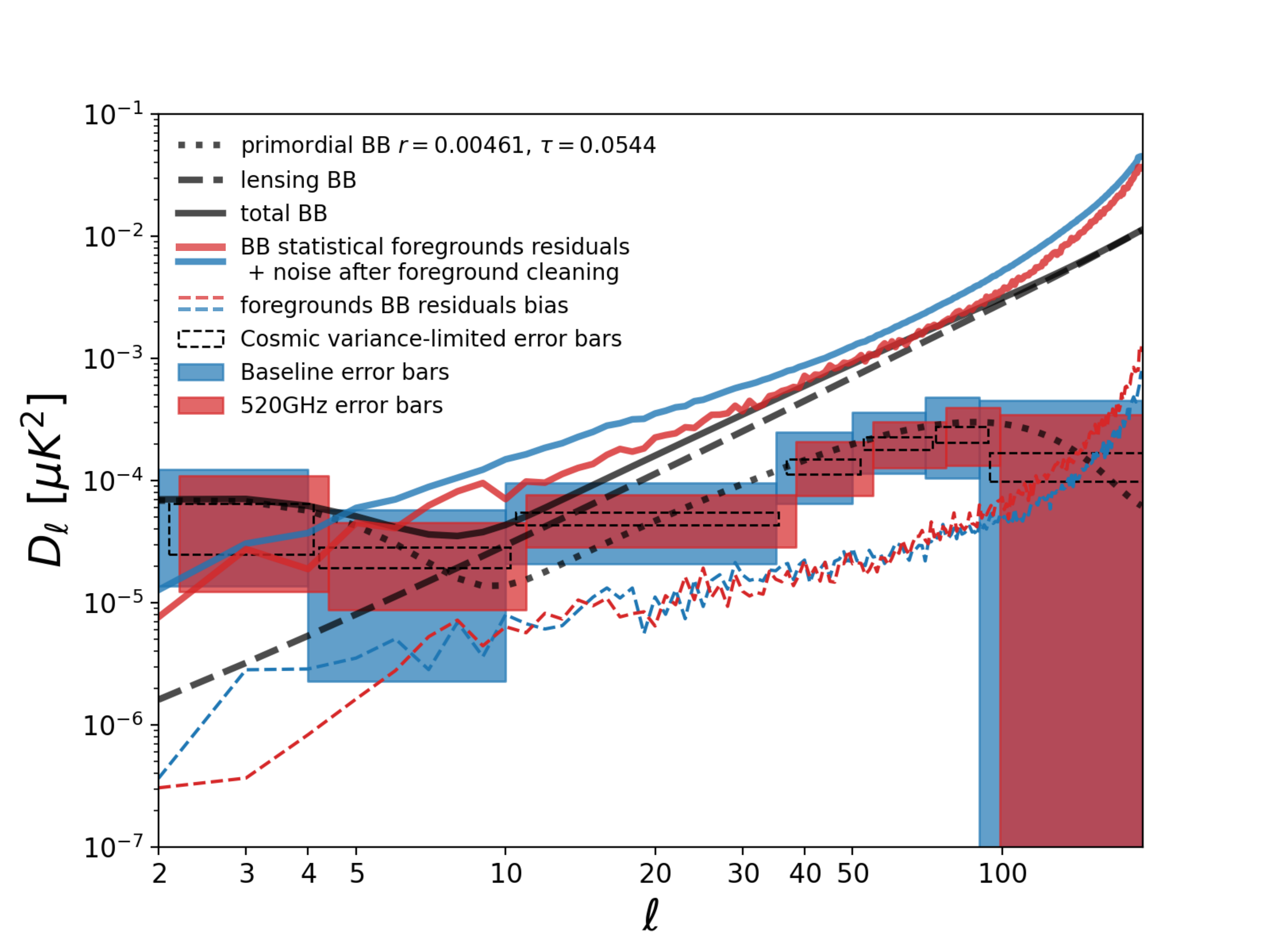}
  \end{center}
  \caption{Example of the results from the \FGBuster\ component-separation runs.
  Two instrumental configurations are considered: the baseline M0 (blue) and the high frequency extension M3 (red).
  The solid lines represent the post-component-separation noise and foreground residuals $S_\ell$, computed by averaging 1000 simulations for the baseline, and averaging 10 simulations for the extension.
  Together with the power from gravitational lensing and primordial $B$-modes, they define the error bars on the CMB power spectrum (boxes).
  }
  \label{fig:fgbuster_considerations}
\end{figure}

We consider the following approximate likelihood for the power spectrum:
\begin{linenomath*} 
\begin{align}
    - 2 \ln \mathcal{L}(r) = f_{\rm sky}\sum_\ell (2 \ell + 1)\left( \frac{D_\ell}{C_\ell} + \ln C_\ell \right) + {\rm const},
    \label{eq:fgb_cl_like}
\end{align}
\end{linenomath*} 
where $f_{\rm sky} = 0.49$, $D_\ell = L_\ell + S_\ell + B_\ell$,  $C_\ell = r T_\ell + L_\ell + S_\ell$, $L_\ell$ is the $BB$ power spectrum from gravitational lensing, and $T_\ell$ is the tensor power spectrum for $r = 1$.
The bias on $r$ is estimated by maximizing this likelihood, while we define $\sigma_r$ to be the Fisher estimate obtained by assuming $B_\ell = 0$. 
We prefer this choice over computing $\sigma_r$ from the width of Eq.~\eqref{eq:fgb_cl_like}, because the latter procedure produces values of $\sigma_r$ that are strongly dependent on the bias, while here we want it to reflect only the statistical constraining power of a given configuration. The various spectra involved in the \FGBuster\ analysis are illustrated in Fig.~\ref{fig:fgbuster_considerations}.

The free parameters of our setup are the $N_{\rm side}$ values of the spectral parameters, and the \LiteBIRD\ instrumental configuration.
Regarding the former, we follow \cite{PTEP}, and split the sky into three (almost) iso-Galactic-latitude regions, each covering approximately 20\,\% of the sky. In each region, every parameter can have different values of $N_{\rm side}$, tuned according to its local signal-to-noise ratio. We define the three regions using \Planck\ post-processing masks adopted for temperature component separation \citep{Planck2018compsep}, the so-called GAL20, GAL40, and GAL60. We find this (arbitrary) choice to significantly help coping with the varying signal-to-noise, but the use of more of these separate regions is certainly possible and could further improve the post-component separation performance -- especially if specialized for dust polarization.
Regarding the choice of the $N_{\rm side}$ values of the spectral
parameter, we find that an accurate estimation of $\beta_\mathrm{d}$
is particularly essential for the \LiteBIRD\ configuration, due to a
small statistical uncertainty but potentially large biases coming from
spatial variability; we use $N_{\rm side}^{\beta_\mathrm{d}} = 64$ in
all the regions. Less accurate estimation of $\beta_\mathrm{s}$ and
$T_\mathrm{d}$ are usually tolerable as was experienced with foreground-only simulations, and following the preliminary results from~\cite{Errard2018}.
For synchrotron emission, we adopt $N_{\rm side}^{\beta_\mathrm{s}} = \left[4,2,2\right]$ for the three regions (counting from lower latitudes), and this is kept constant for all instrumental configurations; we do not expect changes at the high frequencies to significantly affect the fit of the low-frequency foreground.\footnote{Note that the formalism does not assume or exploit possible correlations between the components. In addition to the very different spectral dependence, this should further mitigate the impact of dust and dust-synchrotron correlations on the synchrotron fit.} In contrast, increasing the HFT frequencies does increase the constraining power for the dust SED, and therefore requires a more precise modeling of the spatial variability of $T_\mathrm{d}$ to keep the systematic residuals under control. The analysis sky mask covers $f_{\rm sky} = 49\,\%$ of the sky as introduced in~\cite{PTEP}. In addition to the Galactic plane, this mask removes regions with high residuals.

\subsection{Moment expansion}
\label{app:appendix_mom}
\label{sec:moment_method}

A third parametric component-separation method is referred to as the ``moment-expansion method,'' which models the complexity of the polarized dust SED using parametric fitting with moment expansion coefficients in harmonic space. This formalism was introduced by \citet{Chluba2017} and generalized to polarization in \citet{Vacher2022b,Vacher2022c} and at the cross-frequency power spectra level in \citet{Mangilli2021}. An application of this formalism for component separation with the \LiteBIRD\ mission at high frequencies can be found in \cite{Vacher2022}. The formalism and the pipeline developed in this last study will be applied identically in the present work.

In true experimental conditions, averaging over different SEDs emitted from distinct regions with different physical properties is unavoidable. They appear along the line of sight in the 3D sky, between different lines of sight, inside the beam of the instrument, or when performing a spherical harmonic decomposition to calculate the angular power spectra
over large regions of the sky. Since SEDs depend of the frequency in a non-linear fashion, summing over them when spectral parameters are varying will lead to distortions that should be understood and modeled. 
The moment-expansion method aims at accounting for SED distortions from a canonical model by introducing new terms coming from a Taylor-inspired expansion of the SED around a pivot value and with respect to its spectral parameters (e.g., $\beta_{\mathrm{d}}$ and $T_{\mathrm{d}}$ for an MBB). One can generalize this expansion at the cross-frequency power spectra level $\mathcal{D}_\ell(\nu_i \times \nu_j)$, providing an analytical model that can be fitted over the foreground signal. 

As discussed in \cite{Vacher2022}, providing a correct model for the distortions due to spatial variations of the dust temperature is critical for a mission like \LiteBIRD, while it could safely be ignored for missions like Simons Observatory \citep{Azzoni2021}, with different frequency coverage and lower sensitivity. If the dust signal in every Galactic pixel is indeed a modified black-body, reaching higher frequencies will ease component separation by breaking degeneracies between the moments in $\beta_{\mathrm{d}}$ and $T_{\mathrm{d}}$, providing a better characterization of the impact of the dust temperature variations on the shape of the power spectra SED.

For the present application, we extract the cross-frequency power spectra of $N_{\rm sim}=500$ simulations, including different realizations of Gaussian noise and CMB lensing, with all the high-frequency configurations from M0 to M5. The only foreground component considered is thermal dust with a single MBB. The simulation maps are produced at $N_{\mathrm{side}}=512$ and masked in order to keep a sky fraction of $f_{\rm sky}=70\,\%$, using a raw mask made from \Planck\ intensity data at 353\,GHz as in \citet{Vacher2022}. Only the nine highest frequencies above 100\,GHz are considered, leading to 13 bands (counting the ones sharing the same frequency) and thus 91 cross-spectra. The spectra are binned up to $\ell=200$ with bins of $\Delta \ell= 10$.
Since we expect it to be the only fitting procedure allowing us to recover an unshifted $r$ posterior for the standard \LiteBIRD\ configuration, we shall only consider the $r\beta$--$T$ fitting scheme, using the moment coefficients up to order 1 in both $\beta_{\mathrm{d}}$ and $T_{\mathrm{d}}$ \citep{Vacher2022}. For each simulation, a $\chi^2$ minimization is done at the cross-frequency power spectra level
\begin{linenomath*} 
\begin{equation}
    \chi^2 = \frac{1}{N_{\rm d.o.f.}}\vec{R}^{\sf T}\tens{M}^{-1}\vec{R},
\label{eq:chi2}
\end{equation}
\end{linenomath*} 
with the residual $\mathcal{R}_\ell (\nu_i \times \nu_j) = \mathcal{D}^{\rm sim}_\ell (\nu_i \times \nu_j) - \mathcal{D}^{\rm model}_\ell(\nu_i \times \nu_j)$ and the covariance matrix: $\tens{M}_{\ell,\ell'}^{i\times j,k\times l} = {\rm cov}\left(\mathcal{D}^{\rm sim}_\ell (\nu_i \times \nu_j),\mathcal{D}^{\rm sim}_{\ell'} (\nu_k \times \nu_l)\right)$, allowing us to extract one best-fit value for the tensor-to-scalar ratio $\hat{r}$ per simulation. A histogram is built out of the $N_{\rm sim}$ best fit values of $\hat{r}$, which gives us the final posteriors, with associated standard deviation $\sigma_{\hat{r}}$ found by fitting a Gaussian curve over it.

\section{Minimum variance component-separation methods}
\label{app:blind}
\label{app:appendix_ilc}

\subsection{Internal linear combinations}
\label{subsec:ilc}

Generally speaking, internal linear combination (ILC) methods perform a weighted linear combination of the data $d_\nu$ across frequencies: 
\begin{linenomath*} 
\begin{align}\label{eq:ilc_est}
\hat{s} = \sum_\nu w_\nu\, d_\nu = \vec{w}^{\sf T}\vec{d}\,,
 \end{align}
 \end{linenomath*} 
which is designed to minimize the overall variance due to foregrounds and noise without altering the CMB signal. More explicitly, the ILC weights ${\vec{w}\equiv \{w_\nu\}}$ are the solution of a constrained minimization problem that can be formulated by the Lagrangian,
\begin{linenomath*} 
\begin{align}\label{eq:lagrangian}
 L\left(\vec{w}, \lambda\right) = \vec{w}^{\sf T} \tens{C}\, \vec{w}  +\lambda\left(1 - \vec{w}^{\sf T}\vec{a}\right)\,,
\end{align}
\end{linenomath*} 
in which the matrix $\tens{C}$ collects the elements $\tens{C}_{\nu\nu'}=\langle d_\nu\, d _{\nu'}\rangle$ of the data covariance matrix for all pairs of frequencies $(\nu,\nu')$ and the vector $\vec{a} \equiv \{a_\nu\}$ collects the frequency spectrum of the CMB component across the channels. The Lagrange multiplier $\lambda$ is used to ensure that the overall variance $\vec{w}^{\sf T} \tens{C}\, \vec{w}$ of the ILC estimate $\hat{s}$ is minimized under the constraint $\sum_\nu w_\nu\,a_\nu = 1$ for the preservation of the CMB signal. The ILC weights are thus given by the saddle point of the Lagrangian defined in Eq.~\eqref{eq:lagrangian}:
\begin{linenomath*} 
\begin{align}\label{eq:ilc}
\vec{w}^{\sf T} = \left(\vec{a}^{\sf T}\tens{C}^{-1}\vec{a}\right)^{-1}\vec{a}^{\sf T}\tens{C}^{-1}\,,
 \end{align}
 \end{linenomath*} 
 and as such they do not rely on any explicit parametric foreground model.

\subsection{NILC}
\label{subsec:nilc}

The Needlet ILC (\nilc; \citealp{Delabrouille2009}) method is a specific version of the ILC approach that performs this optimization using a \emph{needlet} (i.e., spherical wavelet) vector basis. In the \nilc\ method, the ILC weights, Eq.~\eqref{eq:ilc}, are computed independently in different regions of the sky and different ranges of angular scales using a wavelet decomposition of the data. Needlets have excellent localization properties, both on the sphere (pixel space) and in harmonic space \citep{Narcowich2006,Guilloux2009}, which allows the \nilc\ algorithm to adjust the ILC weights depending on the local variations of the foreground and noise contaminations across the sky and across different scales. 

\subsection{cMILC}
\label{subsec:cmilc}

The Constrained Moment ILC (\cmilc; \citealp{Remazeilles2021}) method is an extension of the \nilc\ method that includes additional nulling constraints to \emph{deproject} spectral moments of the foreground emission arising from line-of-sight integration and beam convolution. It combines the so-called constrained ILC approach \citep{Remazeilles2011} and the moment expansion technique for dealing with foreground emission \citep{Chluba2017}. Because of spectral variations of foreground emission along the line-of-sight and/or inside the beam, the line-of-sight integration and beam convolution effects in sky observations  cause spectral distortions to the baseline SEDs of thermal dust and synchrotron, and thus induce partial decorrelation across frequencies. Such higher-order corrections to the baseline foreground emission can be significant compared to the CMB $B$-mode signal at low tensor-to-scalar ratio values.

These effects are addressed by \cmilc\  through moment expansion of the foreground emission $I_\nu(\hat{n})$ beyond the leading-order SED:
\begin{linenomath*} 
\begin{align}\label{eq:moment}
I_\nu(\hat{n}) = & \,A(\hat{n})\,f\left(\nu,\overline{\vec{\beta}}\right)\, +\, \sum_i A(\hat{n})\left(\beta_i(\hat{n}) -  \overline{\beta}_i\right){\partial\over \partial\overline{\beta}_i} f\left(\nu,\overline{\vec{\beta}}\right)\cr
                      & + {1\over 2!}\sum_{i,j} A(\hat{n})\left(\beta_i(\hat{n}) -  \overline{\beta}_i\right)\left(\beta_j(\hat{n}) -  \overline{\beta}_j\right){\partial^2\over \partial\overline{\beta}_i\partial\overline{\beta}_j} f\left(\nu,\overline{\vec{\beta}}\right)\cr
                      & + \cdots\,,
\end{align}
\end{linenomath*} 
where $f(\nu,\overline{\vec{\beta}})$ is the baseline SED evaluated at fixed pivot values ${\overline{\vec{\beta}}\equiv \{\overline{\beta}_i\}}$ for a set of spectral parameters ${\vec{\beta}\equiv \{\beta_i(\hat{n})\}}$ (e.g., dust spectral index and temperature) that vary with the direction $\hat{n}$ on the sky and along the line of sight. 
The expansion defined by Eq.~\eqref{eq:moment} highlights the foreground moment components,
\begin{linenomath*} 
\begin{align}
m_{\vec{\bf \alpha}}(\hat{n}) = A(\hat{n})\,{\left(\beta_1(\hat{n}) -  \overline{\beta}_1\right)^{\alpha_1}\cdots\left(\beta_n(\hat{n}) -  \overline{\beta}_n\right)^{\alpha_n} \over \alpha_1!\cdots \alpha_n!},
\end{align}
\end{linenomath*} 
which have homogeneous SEDs given by the derivatives of the baseline SED with respect to spectral parameters:
\begin{linenomath*} 
\begin{align}\label{eq:moment_sed}
b_{\vec{\alpha}}(\nu) = {\partial^{\alpha_1}  \over \partial\overline{\beta}_1^{\,\alpha_1}} \cdots {\partial^{\alpha_n}  \over \partial\overline{\beta}_n^{\,\alpha_n}} \, f\left(\nu,\overline{\vec{\beta}}\right)\,,
\end{align}
\end{linenomath*} 
where $\vec{\alpha} = (\alpha_1,\cdots,\alpha_n) \in \mathbb{N}^n$.

\cmilc\ enables us to \emph{deproject} several moments of the foreground emission from the recovered CMB $B$-mode map by generalizing the Lagrangian in Eq.~\eqref{eq:lagrangian} as
\begin{linenomath*} 
\begin{align}\label{eq:lagrangian2} 
 L\left(\vec{w}, \lambda, \{\mu_{\vec{\alpha}}\}\right) = \vec{w}^{\sf T} \tens{C}\, \vec{w}  +\lambda\left(1 - \vec{w}^{\sf T}\vec{a}\right) + \sum_{\vec{\alpha}}\, \mu_{\vec{\alpha}}\, \vec{w}^{\sf T}\vec{b}_{\vec{\alpha}}\,,
\end{align}
\end{linenomath*} 
in which extra Lagrange multipliers $\mu_{\vec{\alpha}}$ are introduced to impose several nulling constraints on the ILC weights, ${\sum_\nu w_\nu\, b_{\vec{\alpha}}(\nu) = 0}$, with respect to the moment SEDs in Eq.~\eqref{eq:moment_sed}.
The \cmilc\ weights are thus given by the saddle point of the extended Lagrangian in Eq.~\eqref{eq:lagrangian2},
\begin{linenomath*} 
\begin{align}\label{eq:cmilc}
\vec{w}^{\sf T} = \vec{e}^{\sf T}\left(\tens{A}^{\sf T}\tens{C}^{-1}\tens{A}\right)^{-1}\tens{A}^{\sf T}\tens{C}^{-1}\,,
 \end{align}
 \end{linenomath*} 
where the mixing matrix ${\tens{A}=(\vec{a}\, \vec{b}_1\, \cdots \, \vec{b}_m)}$ contains the CMB SED vector $\vec{a}$ in the first column, the moment SED vectors, $\vec{b}_1, \cdots, \vec{b}_m$, as defined by Eq.~\eqref{eq:moment_sed} in the other columns, and ${\vec{e}^{\sf T}=(1\, 0\, \cdots\, 0)}$. The unconstrained foregrounds (i.e., high-order moments), which are not deprojected by \cmilc, are simply mitigated through blind variance minimization, as in the \nilc\ algorithm.

For both \nilc\ and \cmilc, we transform the Stokes $Q$ and $U$ full-sky simulations defined in Sect.~\ref{sec:simulations} into full-sky $B$-mode maps before computing the fits described above, resulting in pure $B$-mode constraints. We then apply these methods to the two most extreme instrument configurations defined in Table~\ref{tab:models}, namely M0 and M5. Foreground cleaning is performed on the full sky for both \nilc\ and \cmilc, while the power spectra and $r$ estimates are computed after masking $f_{\rm sky}=50\,\%$ of the \nilc\ and \cmilc\ $B$-mode maps, where the mask is obtained by thresholding the observed $402$\,GHz $B$-mode map until $50$\,\% of the sky is masked.

The $r$ estimates are derived from the angular power spectrum of the \nilc/\cmilc\ $B$-mode map using the Gaussian likelihood
\begin{linenomath*} 
\begin{align}\label{eq:lkl_ilc}
-2\ln\mathcal{L}(r) = \sum_\ell \frac{\left(\widehat{C}_\ell^{BB} - C_\ell\left(r,A_{\rm L} \right) - \widehat{N}_\ell\right)^2 }{\widehat{\Xi}_\ell^{\,2}}\,, 
 \end{align}
 \end{linenomath*} 
where $\widehat{C}_\ell^{BB}$ is the $B$-mode power spectrum of the reconstructed \nilc/\cmilc\ map, $\widehat{N}_\ell$ is the post-component-separation noise, and $C_\ell\left(r,A_{\rm L} \right)$ is the theoretical $B$-mode power spectrum as a function of $r$ that we fit to the data, with lensing amplitude fixed to $A_{\rm L}=1$ or $0$, depending on either no delensing or full delensing assumptions. The covariance matrix is given by
\begin{linenomath*} 
\begin{align}\label{eq:lkl_cov_ilc}
\widehat{\Xi}_\ell = \sqrt{\frac{2}{\left(2\ell+1\right)f_{\rm sky}}}\,\widehat{C}_\ell^{BB}\,, 
 \end{align}
 \end{linenomath*} 
 which implicitly includes the sample variance of the residual foregrounds and noise that are left in the reconstructed \nilc/\cmilc\ map.

\end{appendix}

\end{document}